\documentclass[manuscript]{acmart}
\usepackage{booktabs} 
\usepackage{color}
\usepackage[ruled]{algorithm2e} 
\usepackage{hyperref}
\setcopyright{none} 
\begin{document}
\title{Composable Probabilistic Inference Networks Using MRAM-based Stochastic Neurons}

\author{Ramtin Zand}
\orcid{0000-0002-1786-1152}
\affiliation{%
  \institution{University of Central Florida}
  \city{Orlando}
  \state{FL}
  \postcode{32816}
  \country{USA}}
\email{ramtinmz@knights.ucf.edu}

\author{Kerem Y. Camsari}
\affiliation{%
  \institution{Purdue University}
  \city{West Lafayette}
  \state{IN}
  \postcode{47906}
  \country{USA}}
\email{kcamsari@purdue.edu}

\author{Supriyo Datta}
\affiliation{%
  \institution{Purdue University}
  \city{West Lafayette}
  \state{IN}
  \postcode{47906}
  \country{USA}}
\email{datta@purdue.edu}
  
\author{Ronald F. DeMara}
\affiliation{%
  \institution{University of Central Florida}
  \city{Orlando}
  \state{FL}
  \postcode{32816}
  \country{USA}}
\email{ronald.demara@ucf.edu}

\begin{abstract}
Magnetoresistive random access memory (MRAM) technologies with thermally unstable nanomagnets are leveraged to develop an intrinsic stochastic neuron as a building block for restricted Boltzmann machines (RBMs) to form deep belief networks (DBNs). The embedded MRAM-based neuron is modeled using precise physics equations. The simulation results exhibit the desired sigmoidal relation between the input voltages and probability of the output state. A probabilistic inference network simulator (PIN-Sim) is developed to realize a circuit-level model of an RBM utilizing resistive crossbar arrays along with differential amplifiers to implement the positive and negative weight values. The PIN-Sim is composed of five main blocks to train a DBN, evaluate its accuracy, and measure its power consumption. The MNIST dataset is leveraged to investigate the energy and accuracy tradeoffs of seven distinct network topologies in SPICE using the 14nm HP-FinFET technology library with the nominal voltage of 0.8V, in which an MRAM-based neuron is used as the activation function.  The software and hardware level simulations indicate that a $784\times200\times10$ topology can achieve less than 5\% error rates with $\sim400 pJ$ energy consumption. The error rates can be reduced to 2.5\% by using a $784\times500\times500\times500\times10$ DBN at the cost of $\sim10\times$ higher energy consumption and significant area overhead. Finally, the effects of specific hardware-level parameters on power dissipation and accuracy tradeoffs are identified via the developed PIN-Sim framework.

\end{abstract}


\maketitle
\thispagestyle{empty}

\section{Introduction}
In recent years, innovation within the disciplines of machine intelligence and learning (ML) utilizing artificial neural networks (ANN) that aim to model biological brain behavior has grown significantly due to the existence of vast datasets available to train such networks. Some interesting projects within these fields include solving complicated classification problems by utilizing ANN's strength in information processing \cite{BASHEER2000}, pattern recognition tasks \cite{bishop1995}, and even out-maneuvering a world champion Go player to a historic defeat \cite{churchland2016}.

The techniques most commonly used to train ANNs today typically require supervised learning, where the error rate is measured by comparing the output from the network with a known desired output. Then, using a subsequent training technique such as backpropagation, the corresponding weights within the network are adjusted \cite{hecht1992}. However, unsupervised learning is attracting considerable attentions in recent years due to its compatibility with the nature of intelligent biological systems, which learn through observation, not by supervision \cite{lecun2015deep}. In unsupervised learning approaches, decision processes based on probabilistic inference are built upon constructing statistical correlation of the inputs into categories \cite{buesing2011}. Deep belief networks (DBNs) are an interesting class of ML techniques utilizing an unsupervised learning approach known as contrastive divergence (CD) \cite{carreira2005}, which demonstrates outstanding learning abilities for various applications such as natural language understanding \cite{Sarikaya2014}. DBNs are constructed by multiple Restricted Boltzmann machines (RBMs), which can be hierarchically connected to form a network \cite{hinton2006}.

Research focused on software implementation of DBNs show that conventional von-Neumann architectures are poorly-matched to the processing flow in terms of the constituent operations at a fine granularity. Although software implementations on conventional architectures provide flexibility, they require significant execution time and energy caused by the memory-processor bandwidth bottleneck, which is intensified due to the large matrix multiplications required \cite{Merolla2014}. Therefore, hardware-based RBM design research seeks to surmount these limitations. Previous work on RBM hardware implementations use conventional VLSI design techniques \cite{Yuan2017}, FPGA approaches \cite{Kim2010,Ly2010}, and stochastic CMOS methods\cite{Ardakani2017}. Moreover, emerging technologies such as resistive RAM (RRAM) \cite{SHERI2015,Bojnordi2016} and phase change memory (PCM) \cite{Eryilmaz2016} had been utilized as weighted connections within the DBN architecture to interconnect its various building blocks. The previous hybrid Memristor/CMOS designs attempt to realize an intrinsic implementation of the weighted connections. Recently, a current-driven low energy-barrier spintronic device has been proposed to be utilized in RBMs as the activation function \cite{zandRDBN}, while similar devices have been previously proposed for spiking \cite{sengupta2016SREP,Sengupta2016TED} and hard axis clocked \cite{behin2016} neural systems. However, the current-mode operation of these devices imposes a significant power consumption to the activation functions, while requiring weighted connections with $M\Omega$ resistances. The design proposed herein takes a new approach from the device-level upward to overcome the challenges mentioned above by utilizing a voltage-driven spintronic device with embedded magnetoresistive random access memory (MRAM) constructed by low energy barrier nanomagnets, which leverages intrinsic thermal noise to provide a natural and power-efficient building block for RBMs. Moreover, we propose a simulation framework for probabilistic learning networks, called PIN-Sim, which is utilized herein to realize a feasible circuit-level implementation of DBN architectures using a SPICE model of our proposed embedded MRAM-based neuron. Specifically, the main contributions of this paper are as follows:

1. A transportable Probabilistic Inference Network Simulator (PIN-Sim) to realize a circuit-level implementation of DBN utilizing voltage-controlled embedded MRAM-based neurons as the probabilistic sigmoidal activation functions. The PIN-Sim framework can be utilized for design space exploration to achieve an optimized network implementation based on the application requirements.

2. Detailed results and analyses about the effects of various circuit-level and device-level tunable parameters on the accuracy and power consumption of the DBNs implemented by PIN-Sim framework.

3. Discussions regarding the effects of noise, and variations in the resistance of the weighted connections on the accuracy of our proposed probabilistic spin logic-based DBN circuits.

The remainder of the paper is organized as follows. Section 2 describes the fundamentals of the RBMs and the CD unsupervised learning algorithm. The structure and modeling methodology of the proposed neuron with embedded MRAM is elaborated in Section 3. Section 4 provides details about the circuit-level implementation of DBNs using our proposed PIN-Sim framework. The software and hardware level simulation results are provided in Section 5, as well as a comprehensive comparison between our proposed DBN realization and previous hardware implementations. Finally, Section 6 concludes the paper by relating its contributions, as well as the improvements achieved by the proposed MRAM-based neuron and PIN-Sim framework.


\begin{figure}
\centering
\includegraphics[scale=0.45]{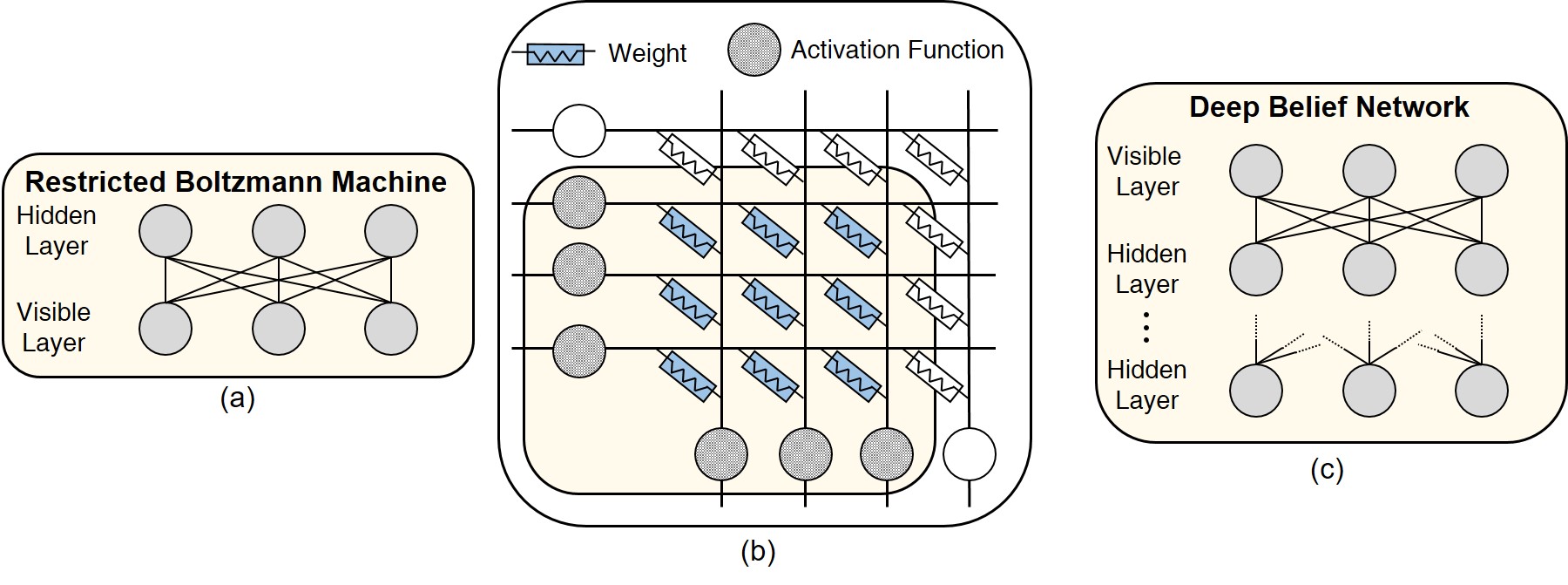}
\caption{(a) An RBM structure depicting neurons organized into hidden and visible layers, (b) a $3\times3$ RBM implemented within a $4\times4$ crossbar architecture using a weighted array to generate resistances needed to appropriately control the activation function, (c) a DBN structure constructed from multiple hidden layers which act to increase recognition accuracy.}
   \label{fig:crossbar}
\end{figure}

\section{Restricted Boltzmann Machines}
Restricted Boltzmann machines (RBMs) are a class of recurrent stochastic neural networks, in which each state of the network, \textit{k}, has an energy determined by the connection weights between nodes and the node bias as described by Equation \ref{eqn:01}, where $s_i^k$ is the state of node \textit{i} in \textit{k}, \textit{b\textsubscript{i}} is the bias, or intrinsic excitability of node \textit{i}, and \textit{w\textsubscript{ij}} is the connection weight between nodes \textit{i} and \textit{j} \cite{ackley1985}.
\begin{equation}
\label{eqn:01}
  E(k) = -\sum_{i} s_i^k b_i -\sum_{i<j} s_i^k s_j^k w_{ij} 
\end{equation}

Each node in an RBM has a probability to be in state one according to Equation \ref{eqn:02}, where $\sigma$ is the sigmoid function. RBMs, when given sufficient time, reach a Boltzmann distribution where the probability of the system being in state \textit{\textbf{s}} is found by Equation \ref{eqn:03}, where \textit{\textbf{u}} could be any possible state of the system. Thus, the system is most likely to be found in states that have the lowest associated energy.

\begin{equation}
\label{eqn:02}
  P(s_i = 1) = \sigma (b_i + \sum_{j} w_{ij} s_j)
\end{equation}

\begin{equation}
\label{eqn:03}
	P(s) = \frac{e^{-E(s)}}{\sum_{u} e^{-E(u)}}
\end{equation}

RBMs are constrained to two fully-connected non-recurrent layers called the \textit{visible layer} and the \textit{hidden layer}. As shown in Figure \ref{fig:crossbar}, RBMs can be readily implemented by a crossbar architecture. The most well-known approach for training RBMs is contrastive divergence (CD), which is an approximate gradient descent procedure using Gibbs sampling \cite{carreira2005}. CD operates in four steps as described below:

1. \textit{Feed-Forward 1:} the training input vector, \textbf{$v$}, is applied to the visible layer, and the hidden layer, \textbf{$h$}, is sampled. 

2. \textit{Feed-back:} The sampled hidden layer output is fed-back and the generated input (\textbf{$v'$}) is sampled. 

3. \textit{Feed-Forward 2:} \textbf{$v'$} is applied to the visible layer and the reconstructed hidden layer is sampled to obtain \textbf{$h'$}. 

4. \textit{Update:} The weights are updated according to Equation \ref{eqn:04}, where $\eta$ is the learning rate and \textbf{$W$} is the weight matrix.

\begin{equation}
\label{eqn:04}
	\Delta W = \eta (vh^T-v'h'^T)
\end{equation}

RBMs can be readily stacked to realize a DBN, which can be trained similarly to RBMs. The training process is conducted by executing CD starting first with the visible layer and the first of the hidden layers within the network. The CD is repeated as many times as required, which will adjust the weights in a hierarchical flow as described in Algorithm \ref{alg:CD}. 

\begin{algorithm}[t]
\SetAlgoNoLine
\KwIn{train dataset $(D_{train})$, \# of training samples $(S)$, \# of RBMs $(M)$}
\KwOut{$weight(n).mat$, $bias(n).mat$, where $n$ is the RBM number}
\textbf{Require:} Maximum iteration $(MaxIter)$, Learning Rate $(\eta)$

\For {i= 1 : S}{
	$\textbf{v}=D_{train}(i)$ \;
    	\For {j=1 : M}{  
    		\For {k=1 : MaxIter}{
        		Feed-Forward 1: $\textbf{h}=\sigma (b + \sum w.\textbf{v})$ \;
                Feed-Back: $\textbf{v'}=\sigma (c + \sum w.\textbf{h})$ \;
                Feed-Forward 2: $\textbf{h'}=\sigma (b + \sum w.\textbf{v'})$ \;
                Update: 
                \\$\Delta \textbf{W}(j) = \eta (vh^T-v'h'^T) \Rightarrow \textbf{W}(j)=\textbf{W}(j)+\Delta \textbf{W}(j)$ \\ $\Delta \textbf{B}(j) = \eta (h-h') \Rightarrow \textbf{B}(j)=\textbf{B}(j)+\Delta \textbf{B}(j)$ \\ $\Delta \textbf{C}(j) = \eta (v-v') \Rightarrow \textbf{C}(j)=\textbf{C}(j)+\Delta \textbf{C}(j)$

        	}
		}
}
\For {j=1 : M}{
	$weight(j).mat \Leftarrow W(j)$ \;
    $bias(j).mat \Leftarrow B(j)$ \;
}

\caption{Contrastive Divergence Unsupervised Learning Algorithm}
\label{alg:CD}
\end{algorithm}

\begin{figure}
\centering
\includegraphics[width=0.8\linewidth]{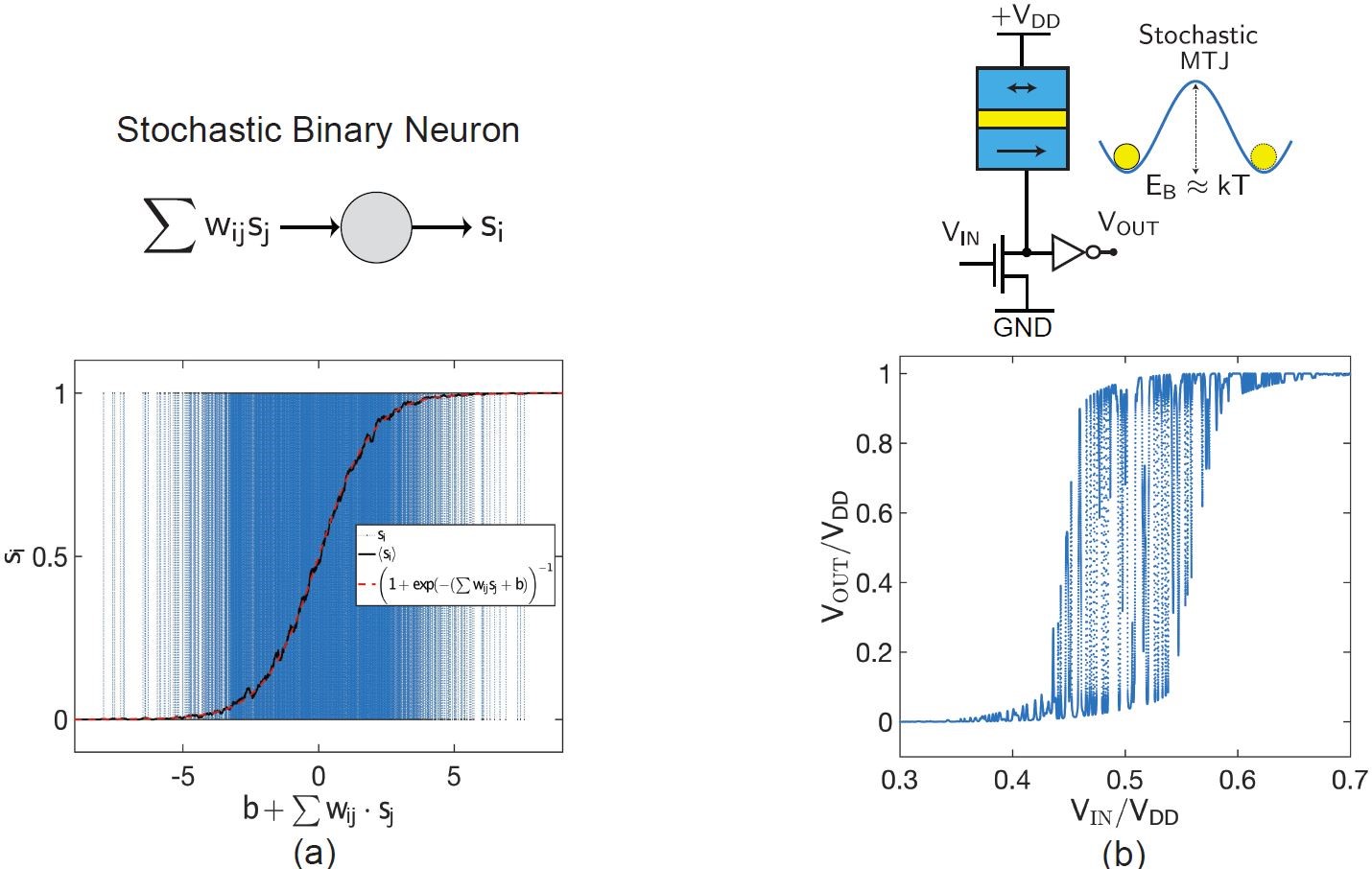}
\caption{a) The building block of the proposed spin-based RBMs, the stochastic binary neuron and its ideal input output characteristics are shown. The dashed red curve indicates the mean of the output that is given by the sigmoid function, $\sf \sigma (z) = 1/(1+exp(-z))$, where z is the input. The dashed blue curve is the instantaneous output while the input is being swept. The running average of the output, as indicated by the black curve, shows a mean that is equal to the sigmoid function. b) A hardware representation of the stochastic binary neuron in terms of an 
Embedded Magnetic Tunnel Junction architecture is shown. The free layer of a conventional Embedded MTJ has an energy barrier $\sf E_B$ of  40-60 kT and thus is non-volatile. Reducing
the energy barrier of the free layer results in a resistive behavior that is  fluctuating between a low ($\sf R_P$ parallel orientation) and a high ($\sf R_{AP}$ anti-parallel) resistance. 
The gate voltage of the transistor ($\sf V_{IN}$) controls the resistance of the transistor to regulate the output voltage to approximate the behavior of a stochastic binary neuron in hardware.}
\label{fig:pbit}
\end{figure} 

\section{Embedded MRAM Based Neuron as a Building Block for RBMs} 
The basic building block of Boltzmann Machines is a stochastic binary neuron that produces a binary output with a given probability. This probability is modulated by the weighted input the neuron receives from the other neurons \cite{hinton1984boltzmann}, as shown Figure~\ref{fig:pbit} (a). Here, we show that a recently proposed building block that leverages the highly scaled embedded magnetoresistive random access memory (MRAM) technology, which is conventionally used as a memory device, can enable  an approximate hardware representation of the binary stochastic neuron in RBM structure as shown in Figure~\ref{fig:pbit} (b). 

The functional component of an MRAM architecture is a magnetic tunnel junction (MTJ) that is a multilayer 2-terminal device that exhibits a resistance change depending on the orientation
of its magnetic layers. One of these magnetic layers is designed to have a fixed magnetic orientation (fixed layer) while the magnetization of the other layer can be switched by a magnetic field or by a spin-polarized current (free layer). In the latter, a current that flows through the fixed layer can exert a ``spin-transfer-torque'' to switch the magnetization of the free layer allowing an electrical writing mechanism \cite{bhatti2017spintronics}. In conventional memory devices, the free layer is designed to have a large energy barrier with respect to the thermal energy (kT) so that the fixed layer can function as a non-volatile memory. In recent years the use of superparamagnetic MTJs that are not thermally stable have been experimentally and theoretically investigated in search of functional spintronic devices \cite{locatelli2014noise,choi2014magnetic,fukushima2014spin, sutton2017intrinsic,debashis2016experimental,liyanagedera2017magnetic,mizrahi2018neural,camsari2017stochastic,zandRDBN}. 

In this paper, we use a recently proposed design that makes minimal modifications to the 1 Transistor / 1 MTJ architecture of the commercially available embedded MRAM technology \cite{camsari2017implementing}. The first modification is to replace the stable free layer with a low-barrier nanomagnet ($E_B \ll 40kT$) that can be achieved by either reducing the total number of spins in the nanomagnet (by reducing $\sf M_s Vol.$, where $\sf M_s$
is the saturation magnetization and Vol. is the volume \cite{bapna2017current})  or by using circular disk magnets that have no preferential easy-axis \cite{debashis2016experimental}. The resistance of an MTJ with such a low-barrier nanomagnet randomly fluctuates between high ($\sf R_{AP}$) and low resistance states ($\sf R_{P}$), creating a fluctuating output voltage at the drain of the NMOS transistor (Figure~\ref{fig:pbit}b). If the transistor resistance that is controlled by the input voltage ($\sf V_{IN}$) is matched to that of the average MTJ resistance at $\sf V_{IN}=V_{DD}/2$, large voltage fluctuations are obtained at the drain output. For typical $\sf {R_{AP}}/R_{P}$ ratios, a CMOS inverter can amplify these fluctuations to produce a rail-to-rail stochastic output at this input value.  Changing the input voltage modulates the transistor resistance, and can suppress these fluctuating outputs either by making the transistor resistance too small and shorting the output to ground, or by making the transistor resistance too high and making the output node $\sf V_{DD}$.  The basic device operation can be understood by considering the MTJ conductance \cite{camsari2017implementing}:
\begin{equation}
G_{MTJ}  = G_0 \left[ 1 + m_z \frac{TMR}{(2+TMR)}\right]
\label{eq:mtj}
\end{equation}
where $m_z$ is the instantaneous free layer magnetization that is fluctuating stochastically in the presence of thermal noise, $G_0$ is the average MTJ conductance, $(G_P + G_{AP})/2$, and $TMR$ is the tunneling magnetoresistance ratio, that is defined as $TMR =(G_P - G_{AP})/G_{AP}$. The voltage division between the transistor and the MTJ (Figure~\ref{fig:pbit}b) produces a drain voltage that can be expressed as: 
\begin{equation}
V_{DRAIN}/V_{DD} = \frac{( 2+TMR)+ TMR \ m_z}{(2+TMR)(1+\alpha)+TMR \ m_z}
\label{eq:drain}
\end{equation}
where we introduce a parameter, $\alpha$, that is defined as the ratio of the transistor conductance ($G_T$) to the average MTJ conductance ($G_0$),  i. e, $\alpha = G_T / G_0$. As the input voltage $\sf V_{IN}$ changes the transistor
conductance $G_T$, the drain output behaves as a noisy inverter. It can be seen from Equation~{\ref{eq:drain}} that the noise amplitude at the drain is maximum when $\alpha \approx 1$, therefore the MTJ resistance is matched to the NMOS resistance ($\alpha=1$) when $\sf V_{IN}/V_{DD}=0.5$ to obtain an output with large fluctuations at the symmetry point.  Even though the drain voltage shows fluctuations of the order of hundreds of mV for typical TMR values, an additional inverter is used to amplify the noise to produce rail-to-rail voltages for a range of input voltages. 

The full circuit behavior of the embedded MRAM based neuron is modeled by a solving the magnetization dynamics of the low barrier nanomagnet using the stochastic Landau-Lifshitz-Gilbert (LLG) equation self-consistently with the transport equations in a SPICE framework \cite{camsari2015modular}. The NMOS transistor is modeled by the predictive technology models (PTM) and for simplicity a bias-independent MTJ model is used that is modeled according to Equation~{\ref{eq:mtj}}. The magnetization input for the MTJ conductance is instantaneously provided from the stochastic LLG equation. The stochastic LLG reads:
\begin{equation}
\label{eq:llg}
(1+\alpha^2){d\hat m}/{dt} = -|\gamma|{\hat m \times \vec{H}} - \alpha |\gamma| (\hat m \times \hat m \times \vec{H})+  {1}/{q  N}(\hat m \times \vec{I}_{S} \times \hat m)  + \left({\alpha}/{q N} (\hat m \times \vec{I}_{S})\right)
\end{equation}

\noindent where $\alpha$ is the damping coefficient of the nanomagnet, $\gamma$ is the electron gyromagnetic ratio, q  is the electron charge, and $\vec{I}_S$ is the spin current incident to the free layer. The spin current is polarized along the direction of the fixed layer polarization ($\hat z$) and its amplitude is proportional to the charge current $I_c$  flowing through the MTJ, such that  $\vec{I}_S = P I_c \hat z$. $N$ is the total number of spins in the free layer (CoFeB), $N=M_s \mathrm{Vol.}/\mu_B$, where $M_s$ is the saturation magnetization of CoFeB and $\mu_B$ is the Bohr magneton. For the free layer, we use a monodomain circular disk magnet whose effective field $\vec{H}$ is given as $-4\pi M_s m_x \hat {x}  + \vec{H}_n$, $\hat x$ being the out-of-plane direction of the magnet. $\vec{H}_n$ is the isotropic thermal noise field, uncorrelated in three directions: $\left(H_n^{x,y,z}\right)^2=2  \alpha  kT / (|\gamma| M_s \mathrm{Vol.})$. The transistors are based on 14nm HP-FinFET PTM \cite{predictive_tech}.

In this paper, we use a circular disk magnet with $\ll kT$ energy barrier in the absence of any shape anisotropy. Such magnets have been fabricated and characterized in \cite{Debashis2018,Cowburn1999,Vaibhav2018}. Moreover, elliptical magnets showing $GHz$ telegraphic oscillations have also been experimentally observed in \cite{Pufall2004}. The demonstrated parameters listed in  Table~\ref{tab:param} \cite{camsari2017implementing} are used to generate all of the results that are provided within this paper. We also note for the chosen parameters with a circular free layer with an in-plane anisotropy that the results are not significantly influenced by the current that is flowing at the midpoint ($\sf V_{IN}=V_{DD}/2)$, and note that any pinning at higher input voltages benefits the switching operation of the device.  

\begin{table}[]
\centering
\small
\caption{Parameters Used for Modeling and Simulation \cite{camsari2017implementing}}
\label{tab:param}
\begin{tabular}{cc}
\hline
\multicolumn{1}{c}{\textbf{Parameters}} & \textbf{Value}      \\ \hline
Saturation magnetization (CoFeB) $(M_s)$   & $1100 emu/cc$ \cite{sankey2008}        \\
Free Layer diameter, thickness                     & $22 nm$, $2 nm$   \\
Polarization                            & $0.59$  \cite{lin2009}              \\
TMR                                     & $110\%$ \cite{lin2009}               \\
MTJ RA-product                          & $9\Omega - \mu m^2$ \cite{lin2009}\\
Damping coefficient                     & $0.01$ \cite{sankey2008}                \\
Temperature                             & $26.85^\circ C$             \\ \hline
\end{tabular}
\end{table}


\begin{figure}
\centering
\includegraphics[scale=0.36]{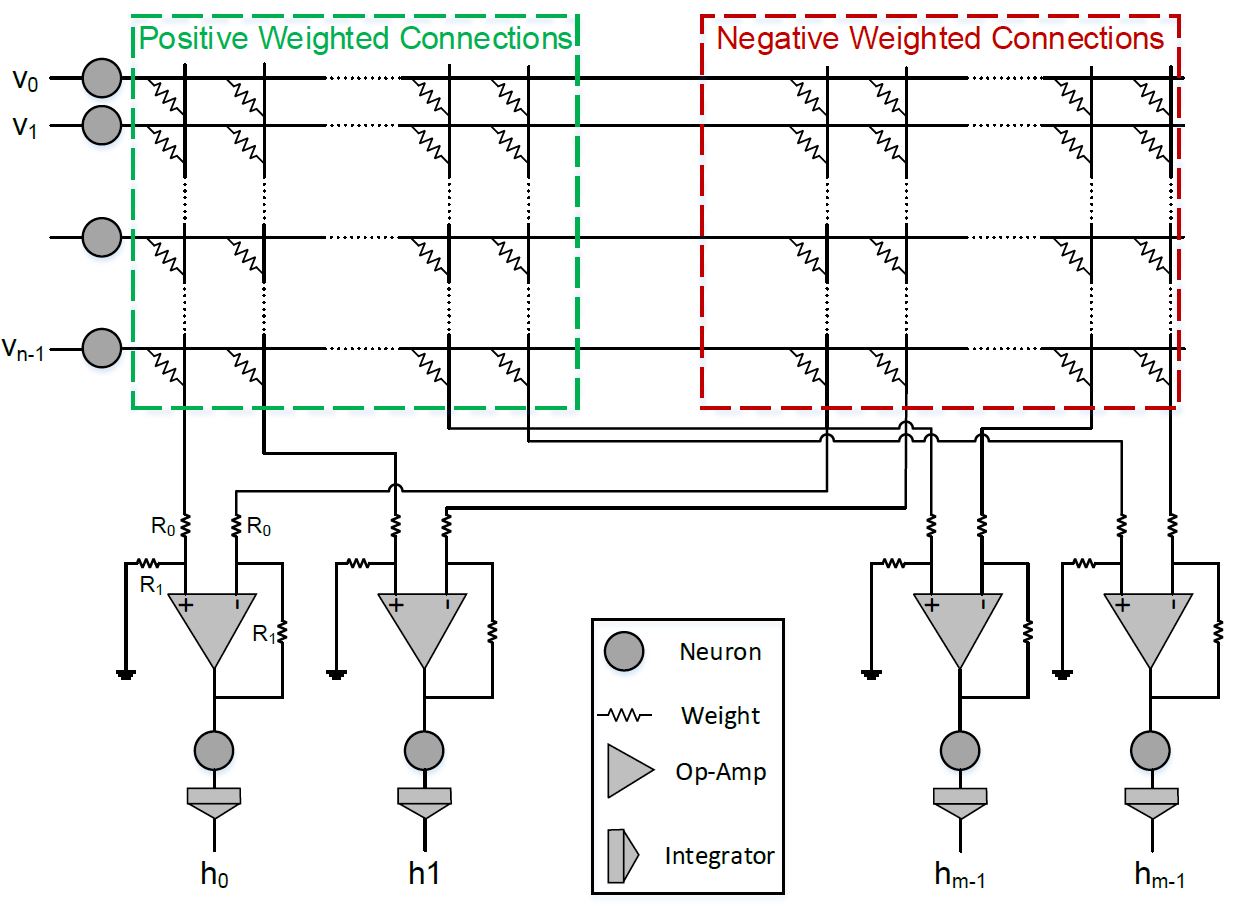}
\caption{An $n \times m$ RBM hardware implementation. Two resistive arrays are leveraged along with differential amplifiers to implement both positive and negative weights. The embedded MRAM-based neurons are used to evaluate the activation functions. The fluctuating output voltage of the neurons are integrated through an RC circuit to generate the output of the proposed RBM structure.}
\label{fig:rbm}
\end{figure}

\begin{figure}
\centering
\includegraphics[scale=0.66]{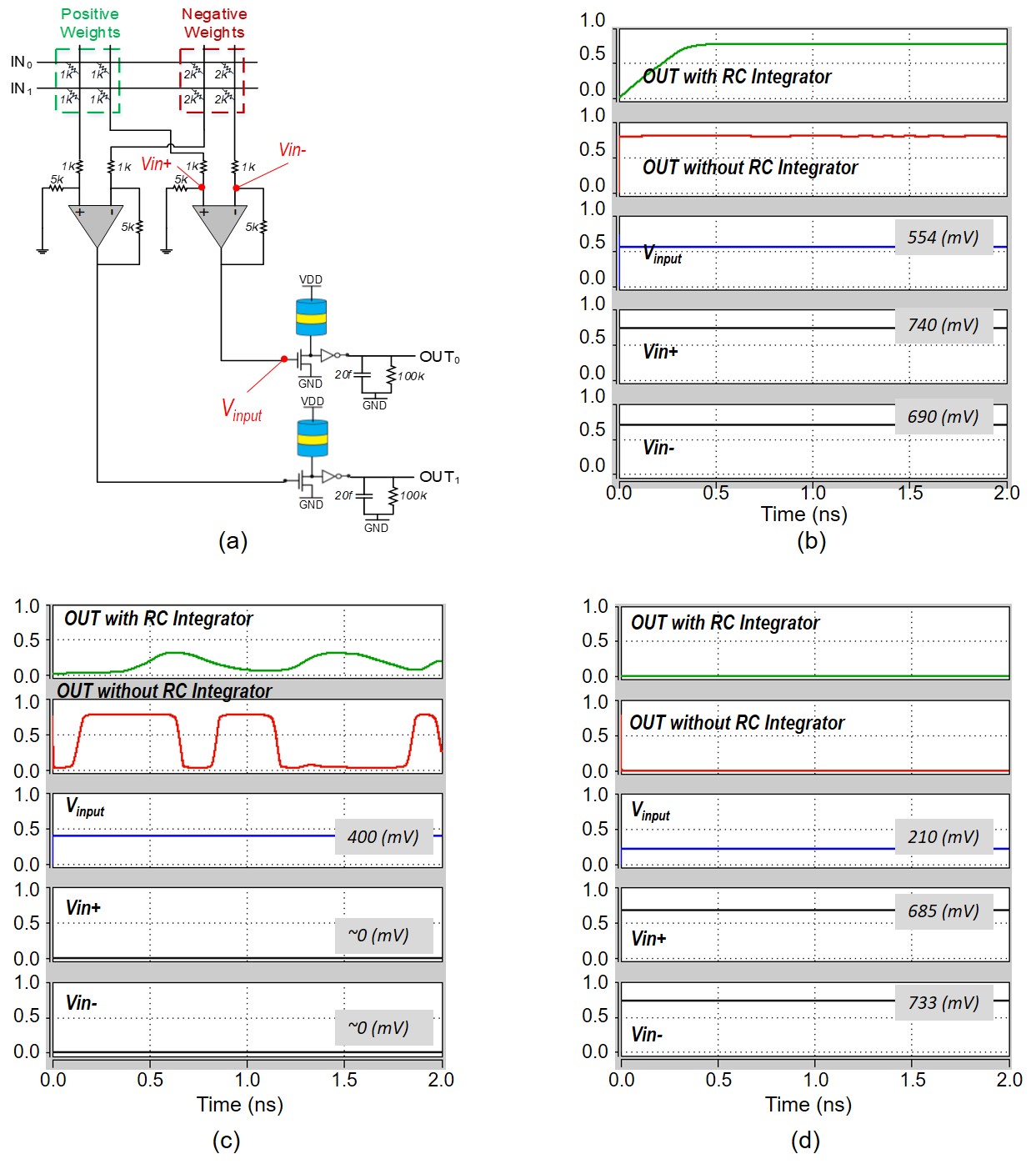}
\caption{(a) a $2\times2$ RBM implementation using the embedded MRAM based neuron. The DC bias voltage of $V_{DD}/2=400 mV$ is added to the output of the differential amplifier to set our proposed neuron at its midpoint. (b) The behavior of the implemented RBM for $IN_{0}=V_{DD}$ and $IN_{1}=V_{DD}$ while the positive and negative weight resistances are $1 k\Omega$ and $2 k\Omega$, respectively. The input voltage connected to the positive terminal of the differential amplifier is larger than the negative terminal resulting in an output voltage larger than VDD/2. The output of the differential amplifier is connected to the input of the neuron, thus the $V_{IN}/V_{DD} = \sim 0.7$ for the neuron leading to output logic ``1'', as shown in Figure \ref{fig:pbit} (b). (c) The behavior of the RBM for $IN_{0}=0$ and $IN_{1}=0$. The inputs of the differential amplifiers are near zero, thus $V_{IN}/V_{DD} = \sim 0.5$ and the state of the neuron fluctuates between ``0'' and ``1''. (d) The RBM behavior for $IN_{0}=V_{DD}$ and $IN_{1}=V_{DD}$ while the positive and negative weight resistances are $2 k\Omega$ and $1 k\Omega$, respectively. The $V_{IN}/V_{DD} = \sim 0.3$ resulting in the neuron being in state ``0'' according to Figure \ref{fig:pbit} (b).}
\label{fig:2by2}
\end{figure}

\subsection{RBM Hardware Implementation}
Figure~\ref{fig:rbm} exhibits a feasible hardware implementation of an $n \times m$ RBM, in which neurons based on the concise embedded MRAM-based design described in the previous section are used to generate the required probabilistic sigmoidal activation function. The resistive crossbar arrays are utilized to realize the matrix multiplication elaborated in Equation~\ref{eqn:02}. In this work, the weights are trained off-chip and the resistive weighted connections will be programmed accordingly. Any resistive devices such as memristors \cite{strukov2008} or spin-orbit torque (SOT)-driven domain wall motion (DWM) devices \cite{Sengupta2016hybrid} can be utilized for weighted connections without the loss of generality.

\section{Proposed DBN structure}

To implement the positive and negative weights in the \textbf{\textit{w}} matrix, two resistive weighted arrays with the same dimensions are required \cite{Hu2012}, as shown in Figure~\ref{fig:rbm}. The outputs of the positive and negative weighted connections are linked to differential amplifiers which are implemented by op-amps as shown in Figure~\ref{fig:rbm}. The output voltage of the op-amp, i.e. $V_{out}=\frac{R_{1}}{R_{0}}(V_{in}^{+}-V_{in}^{-})$, is applied to the MRAM-based neuron as an input signal. The neuron with embedded MRAM will generate an output voltage signal, which fluctuates between VDD and GND with a probability that is modulated based on the applied input voltage. Finally, a resistor-capacitor (RC) integrator circuit is utilized to convert the probabilistic output of the neuron to an analog voltage level, which can be later converted to a digital output through digital to analog conversion. In order to verify the functionality and assess the performance of our proposed RBM implementation, we have simulated a $2\times2$ RBM via SPICE circuit simulation using the 14nm HP-FinFET technology library with an MRAM-based neuron used as the activation function. The results obtained validate the functionality of our proposed design as elaborated in Figure \ref{fig:2by2}.

\begin{figure}
\centering
\includegraphics[scale=0.38]{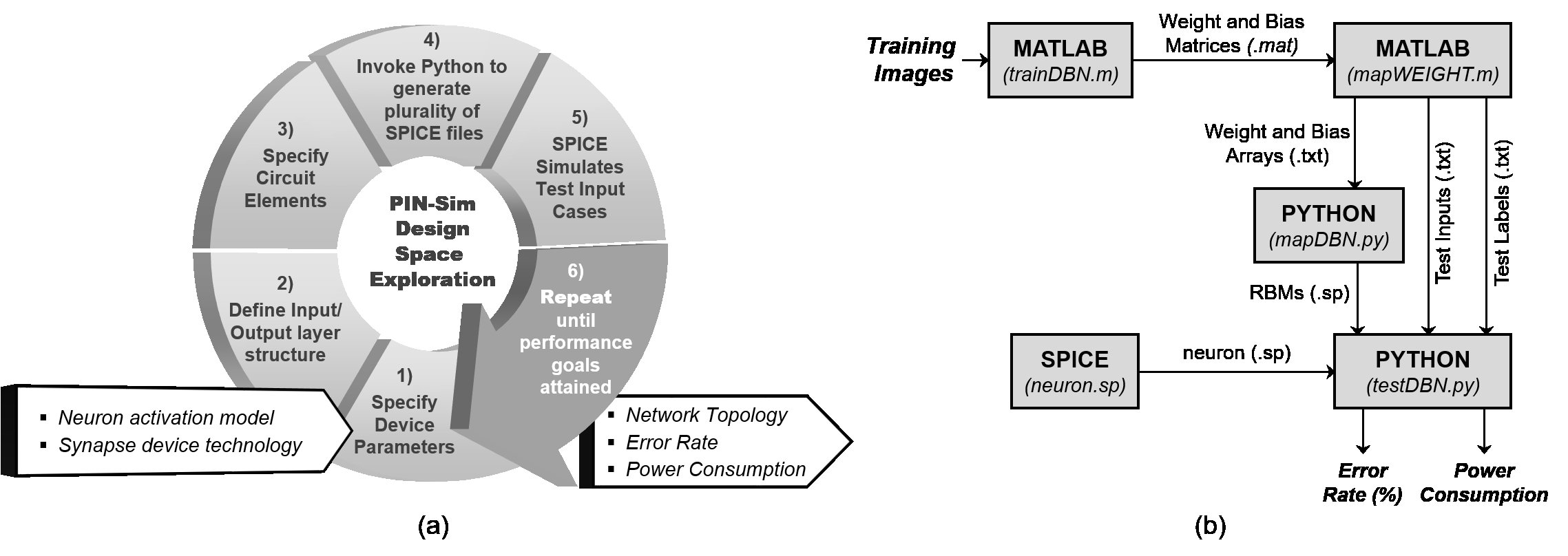}
\caption{(a) The PIN-Sim framework can be utilized to explore the design space to realize the optimized network implementation based on the application requirements. (b) The block diagram of the PIN-Sim framework, which consists of five main modules: \textit{(1) trainDBN:} a MATLAB-based module used for training the DBN architecture. \textit{(2)mapWeight:} a module developed in MATLAB that converts the trained weights and biases to their corresponding resistance values. \textit{(3) mapDBN:} a Python-based module which provides a circuit-level implementation of the RBMs using the obtained weight and bias resistances. \textit{(4) neuron:} A SPICE model of the MRAM-based stochastic neuron. \textit{(5) testDBN:} the main module developed in Python that executes test evaluations to assess the error rate and power consumption using the outputs of the other modules in PIN-Sim.}
\label{fig:pinsim}
\end{figure}

\begin{algorithm}[t]
\SetAlgoNoLine
\KwIn{test dataset $(D_{test})$ with the target labels $(Label)$, \# of test samples$(S)$, \#of RBMs$(M)$,\\ \#of nodes in hidden layer $x$ $(N_x)$}
\KwOut{Error Rate}

\textbf{Initialize:} $Err=0$


$weight.mat, bias.mat \Leftarrow$ \textbf{Contrastive\_Divergence} Algorithm

$posWeight.txt, negWeight.txt, posBias.txt, negBias.txt \Leftarrow \textbf{mapWeight}(Weight.mat, Bias.mat)$ 

\For{i= 1 : S}{
	$input\_data = D_{test}(i)$ \;
	\For{j= 1 : M}{
        $RBM(j).sp \Leftarrow \textbf{mapRBM}(input\_data, N_{j+1}, posWeight.txt, negWeight.txt, posBias.txt, negBias.txt) $\;
        
        Run $RBM(j).sp$ in HSPICE and store the obtained output voltages in array $outRBM$\;
        	\For{k= 1 : $N_{j}$}{
            	Run $neuron.sp$ model with $outRBM(k)$ as the input of the $k_{th}$ Neuron\;           		}
            
            Store the output of the neurons in array $OUTPUT$ \; 
            
            \eIf{( j = M )}{
            	\If {(OUTPUT $\neq$ Label(i))}{
                	$Err +=1$ \;
                	}
            }
        	{
            $input\_data = OUTPUT$ \;
            }

    	}
}

$Error Rate = Err/S$ \;
\caption{PIN-Sim Methodology}
\label{alg:PIN-Sim}
\end{algorithm}

\begin{algorithm}[t]
\SetAlgoNoLine
\KwIn{$weight.mat, bias.mat$, \#of RBMs $(M)$}
\KwOut{$posWeight(n).txt, negWeight(n).txt, posBias(n).txt, negBias(n).txt$, where $n$ is the RBM number}

\textbf{Require:} $r_{min}, r_{max}$, Quantization Factor $(Q)$

$g_{max}=1/r_{min}$\;
$g_{min}=1/r_{max}$\;
$Q=Q/(r_{max}-r_{min})$

\For{i= 1 : M}{
	$W^{+} , W^{-} \Leftarrow weight(i)$ Matrix \;
    
    $B^{+} , B^{-} \Leftarrow bias(i)$ Matrix \;
    
    $w_{min}$ = smallest weight value in $W_{pos} , W_{neg}$  \;
    
    $w_{max}$ = largest weight value in $W_{pos} , W_{neg}$ \;
    
    $b_{min}$ = smallest weight value in $B_{pos} , B_{neg}$  \;
    
    $b_{max}$ = largest weight value in $B_{pos} , B_{neg}$ \;
    
    $GW^{+}=\frac{(g_{max}-g_{min}) \times (W^{+} - w_{min})}{w_{max}-w_{min}} + g_{min}$	,	 $RW^{+}=\frac{round (Q \times 1/GW^{+})}{Q}$\;
    
    
    $GW^{-}=\frac{(g_{max}-g_{min}) \times (W^{-} - w_{min})}{w_{max}-w_{min}} + g_{min}$ , $RW^{-}=\frac{round (Q \times 1/GW^{-})}{Q}$\;
    
    $GB^{+}=\frac{(g_{max}-g_{min}) \times (B^{+} - b_{min})}{b_{max}-b_{min}} + g_{min}$ , $RB^{+}=\frac{round (Q \times 1/GB^{+})}{Q}$\;
    
    $GB^{-}=\frac{(g_{max}-g_{min}) \times (B^{-} - b_{min})}{b_{max}-b_{min}} + g_{min}$ , $RB^{-}=\frac{round (Q \times 1/GB^{-})}{Q}$\;
    
    $posWeight(i).txt \Leftarrow RW^{+}$ \;
    $negWeight(i).txt \Leftarrow RW^{-}$ \;
    $posBias(i).txt \Leftarrow RB^{+}$ \;
    $negBias(i).txt \Leftarrow RB^{-}$ \;
}
\caption{mapWeight Methodology}
\label{alg:mapweight}
\end{algorithm}

\subsection{Probabilistic Inference Network Simulator (PIN-Sim)}
In order to automate and scale up the design space exploration of DBNs at the circuit-level, we have developed a hierarchical simulation framework called PIN-Sim, which can be utilized to implement any probabilistic learning networks. The block diagram of the PIN-Sim framework used to implement DBNs in our work is shown in Figure~\ref{fig:pinsim}, which is comprised of five primary blocks. The PIN-Sim methodology is described in Algorithm~\ref{alg:PIN-Sim}. First, we have modified a MATLAB implementation of DBN developed in \cite{Tanaka2014} to train the network and obtain the trained weight (\textbf{\textit{W}}) and bias (\textbf{\textit{B}}) matrices according to Algorithm~\ref{alg:CD}. The extracted (\textbf{\textit{W}}) and (\textbf{\textit{B}}) matrices are then applied to a MATLAB module called \textbf{\textit{mapWEIGHT}}, the functionality of which is described in Algorithm~\ref{alg:mapweight}. The \textit{mapWEIGHT} module first converts each of the \textbf{\textit{W}} and \textbf{\textit{B}} matrices with positive and negative elements to two separate matrices with only positive elements as described below:

\begin{equation}
	w_{(i,j)}^{+} = \begin{cases} w_{(i,j)}, & \mbox{if } w_{(i,j)} \geq 0  \\ 0, & \mbox{if } w_{(i,j)} < 0 \end{cases}, \> \> \> \> \> \> \> \> \> \> \> \>    
    w_{(i,j)}^{-} = \begin{cases} 0, & \mbox{if } w_{(i,j)} \geq 0  \\ -w_{(i,j)}, & \mbox{if } w_{(i,j)} < 0 \end{cases}
\end{equation}

\begin{equation}
	b_{j}^{+} = \begin{cases} b_{j}, & \mbox{if } b_{j} \geq 0  \\ 0, & \mbox{if } b_{j} < 0 \end{cases}, \> \> \> \> \> \> \> \> \> \> \> \>    
    b_{j}^{-} = \begin{cases} 0, & \mbox{if } b_{j} \geq 0  \\ -b_{j}, & \mbox{if } w_{j} < 0 \end{cases}
\end{equation}

Next, the \textit{mapWEIGHT} module maps the elements in $W^{+}$, $W^{-}$, $B^{+}$, and $B^{-}$ matrices to their corresponding conductance values using the below equations:   

\begin{equation}
\label{eqn:mapw1}
\forall w_{(i,j)} \in (W^{+}, W^{-}): gw_{(i,j)}=\frac{(g_{max}-g_{min}) \times (w_{(i,j)} - w_{min})}{w_{max}-w_{min}} + g_{min}   
\end{equation}
\begin{equation}
\label{eqn:mapw2}
\forall b_{(i,j)} \in (B^{+}, B^{-}): gb_{(i,j)}=\frac{(g_{max}-g_{min}) \times (b_{(i,j)} - b_{min})}{b_{max}-b_{min}} + g_{min}   
\end{equation}
where $\forall g_{(i,j)} \in \textbf{G}: g_{min} \leq g_{(i,j)} \leq g_{max}$, in which $g_{min}=1/r_{max}$ and $g_{max}=1/r_{min}$ are minimum and maximum conductances of all weighted connections in the crossbar weighted array. Moreover, $b_{max}$, $b_{min}$, $w_{max}$, and $w_{min}$ are the maximum and minimum values in all of the bias and weight matrices, respectively. Finally, Equation~\ref{eqn:quant} is utilized to convert and quantize all of the obtained conductance values to their corresponding resistance values, which can then be utilized to implement the required resistive crossbar array.

\begin{equation}
\label{eqn:quant}
\forall g_{(i,j)} \in (GW^{+}, GW^{-}, GB^{+}, GB^{-}): r_{(i,j)}=\frac{round (Q \times 1/g_{(i,j)})}{Q}    
\end{equation}
where $Q$ is the quantization factor, and $GW^{+}$, $GW^{-}$, $GB^{+}$, and $GB^{-}$ are positive weight, negative weight, positive bias, and negative bias conductance matrices, respectively.

Once the positive and negative weight and bias resistance matrices are obtained, they will be converted to \textit{text} files and applied to a Python module called \textbf{\textit{mapRBM.py}}, shown in Figure~\ref{fig:pinsim}, which produces plural crossbar weighted array circuits in SPICE automatically based on the defined network topology. Finally, a \textit{\textbf{testDBN.py}} module is developed using Python scripts, which utilize the generated circuit of the DBN, and the model of the probabilistic neuron to perform a SPICE circuit simulation and calculate the error rate using the \textit{test inputs} and \textit{test labels}, which are provided for the \textit{testDBN} module in form of \textit{text} files.

\section{Simulation Results and Discussion}
Herein, we have leveraged a hierarchical simulation method to examine the performance of our DBN implementation. In software-level simulation, the behavioral results of the developed embedded MRAM-based neuron model are used to implement a DBN in MATLAB for MNIST pattern recognition application \cite{Lecun1998}. In the hardware-level simulation, the proposed framework is used to develop a circuit-level DBN implementation using the p-bit SPICE model and 14nm CMOS technology in SPICE circuit simulator with 0.8V nominal voltage.

\begin{figure}
\centering
\includegraphics[scale=0.6]{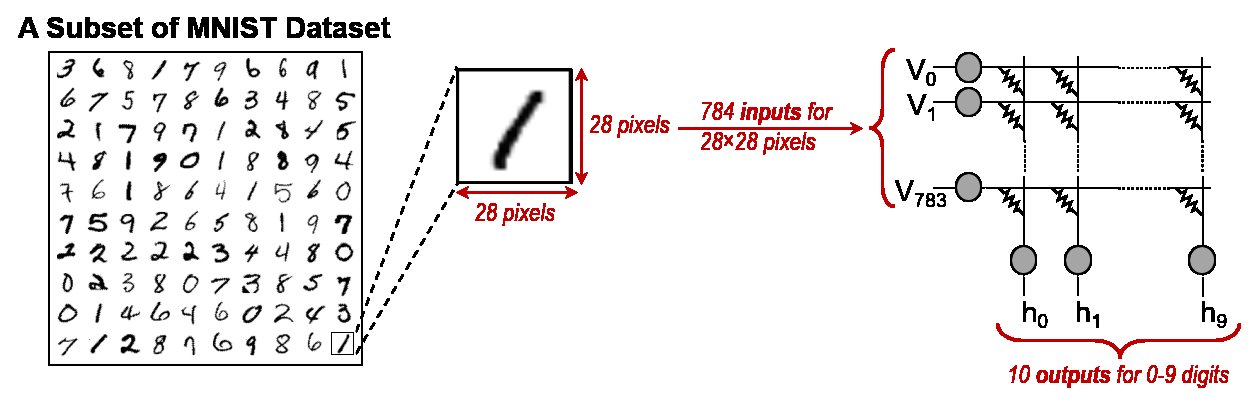}
\caption{The most elementary $784 \times 10$ DBN required for MNIST digit recognition application. The visible layer includes 784 nodes to handle $28 \times 28$ pixels of the input images, while the 10 nodes in hidden layer represent the output classes. }
\label{fig:MNIST}
\end{figure}
\subsection{MATLAB simulation}

\begin{figure}
\centering
\includegraphics[scale=0.3]{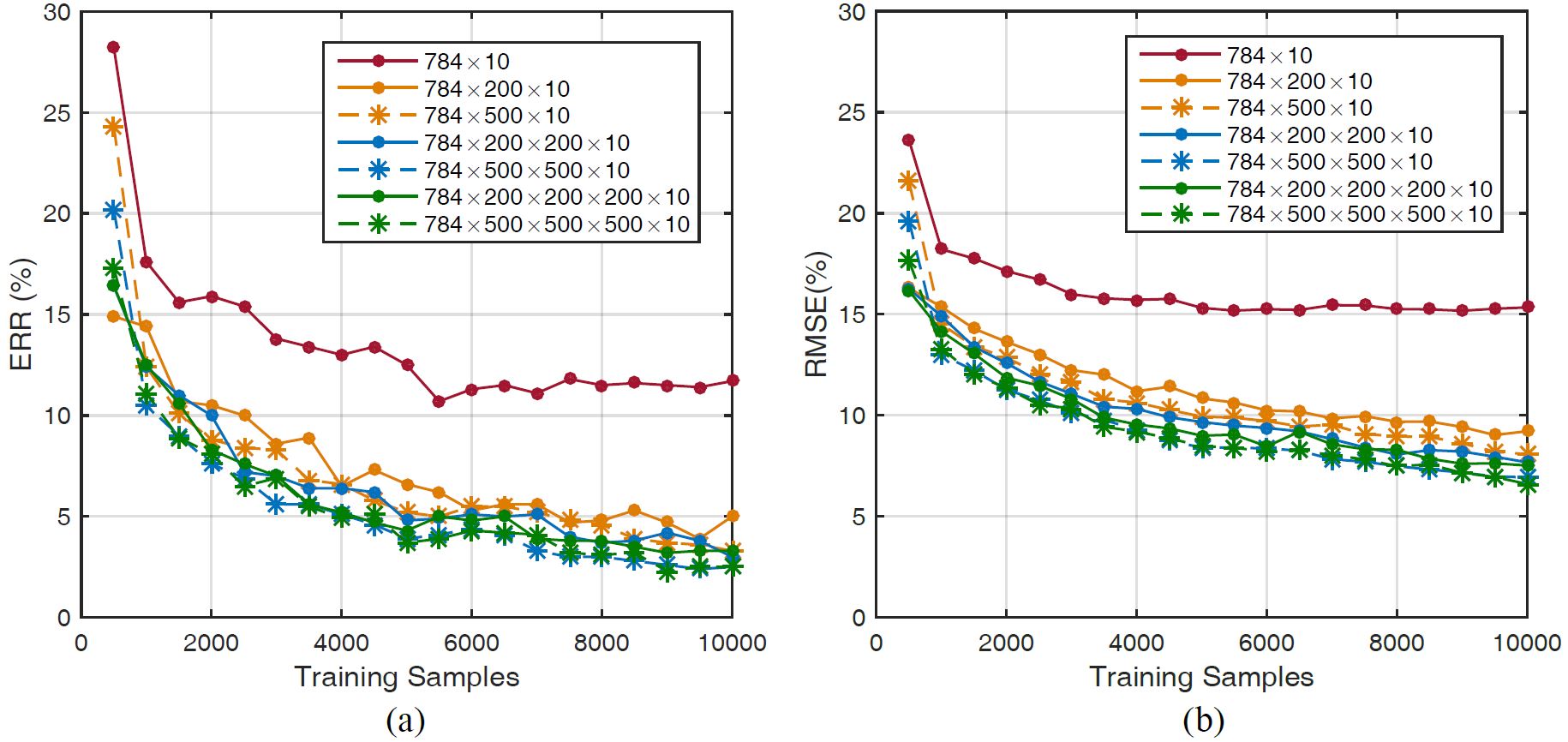}
\caption{(a) ERR vs. training samples for various DBN topologies, (b) RMSE vs. training samples for various DBN topologies.}
\label{fig:err}
\end{figure}

Herein, we have modified the sigmoid activation function in a MATLAB implementation of DBN \cite{Tanaka2014} by using the device-level simulation results of the proposed embedded MRAM-based neuron. To assess the performance of the implemented DBN, we have used the MNIST data set \cite{Lecun1998} including 60,000 training and 10,000 test sample images of hand-written digits, each of which having $28 \times 28$ pixels. We have used Error rate (ERR) and root-mean-square error (RMSE) metrics to evaluate the performance of the DBN, as expressed by the following equations \cite{Tanaka2014}:

\begin{equation}
\label{eqn:err}
ERR=\frac{N_F}{N}    
\end{equation}

\begin{equation}
\label{eqn:rmse}
RMSE=\sqrt[]{\frac{1}{MN} \sum_{k=1}^N (y_k-F(x_k)^2)}  
\end{equation}
where $M$ is the number of output classes, $N$ is the number of input data, $N_F$ is the number of false inference, $F$ is the inference of the trained DBN, $x_k$ is the $k$-th input data and $y_k$ represents its corresponding target output.

As shown in Figure~\ref{fig:MNIST}, the most elementary model of the DBN requires 784 nodes in visible layer for the $28 \times 28$ pixels of the input images, and 10 nodes in hidden layer for, which represents 0-9 output digits. Figure~\ref{fig:err} shows the relation between the error rate and the number of training samples for seven distinct DBN topologies, which is obtained using 1,000 test samples. The results obtained by MATLAB simulation exhibit that an error rate of 28.2\% for a $784 \times 10$ DBN trained by 500 training inputs can be decreased to a 2.5\% error rate achieved using $784 \times 500 \times 500 \times 500 \times 10$ and $784 \times 500 \times 500 \times 10$ DBN topologies, which are trained by 10,000 input training samples. Thus, the recognition accuracy can be improved by increasing the number of hidden layers in the network, number of nodes in each layer, and number of training samples. However, these improvement can lead to higher power consumption and area overheads as investigated in the hardware-level simulations elaborated below.

\begin{figure}
\centering
\includegraphics[scale=0.4]{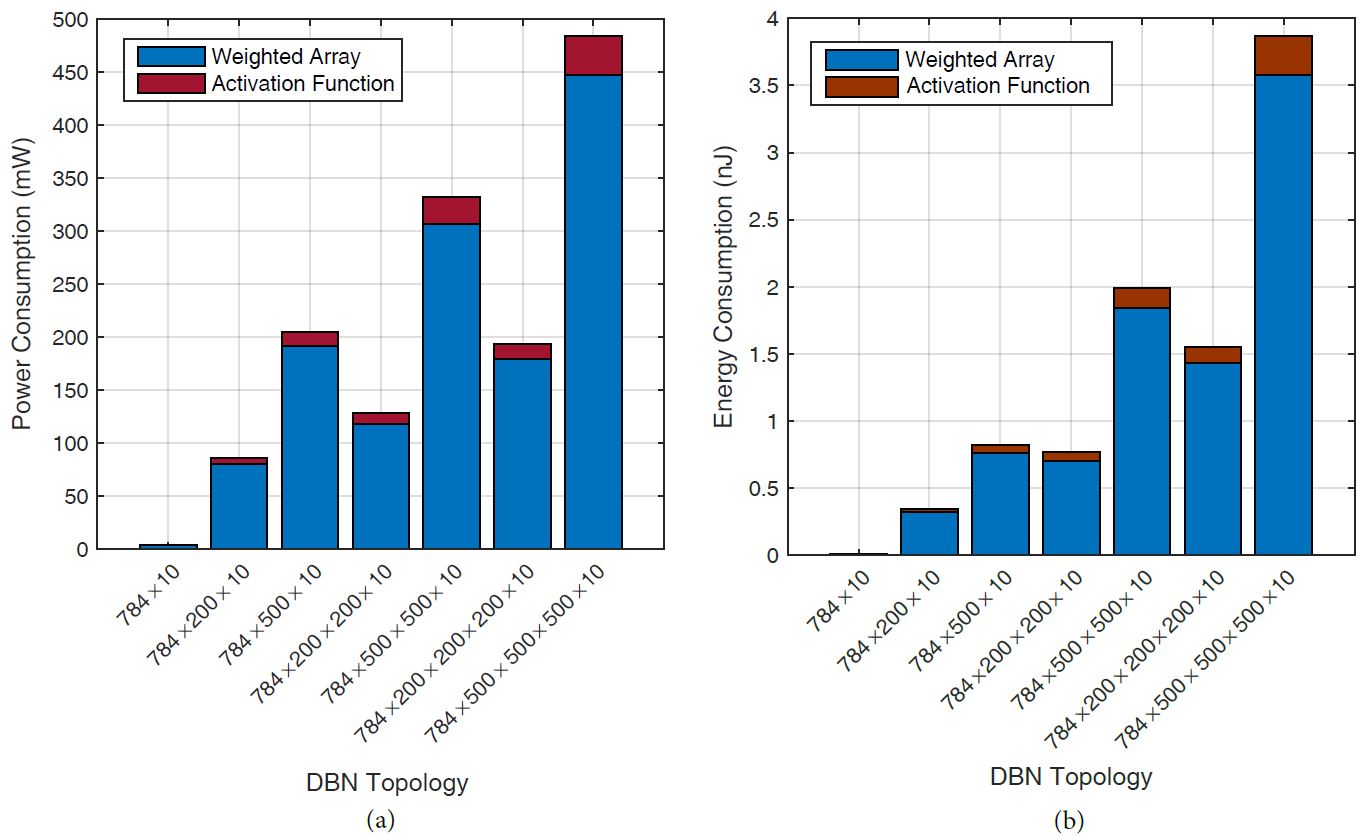}
\caption{Test operation: (a) Power Consumption for various DBN topologies, (b) Energy Consumption for various DBN topologies.}
\label{fig:pwr-eng}
\end{figure}

\subsection{PIN-Sim simulation}
In this section, we utilize our proposed PIN-Sim framework to provide a circuit-level model of DBN architecture. Next, we will provide the energy and power consumption profiles of the seven different DBN topologies investigated in the previous section to analyze the energy and accuracy trade-offs of these networks. Finally, we will focus on the effect of various important hardware-level parameters. These are vital parameters during design space exploration that influence the accuracy of DBN architectures as tradeoffs necessary to obtain efficient hardware-level implementation for pattern recognition applications.

\begin{figure}
\centering
\includegraphics[scale=0.43]{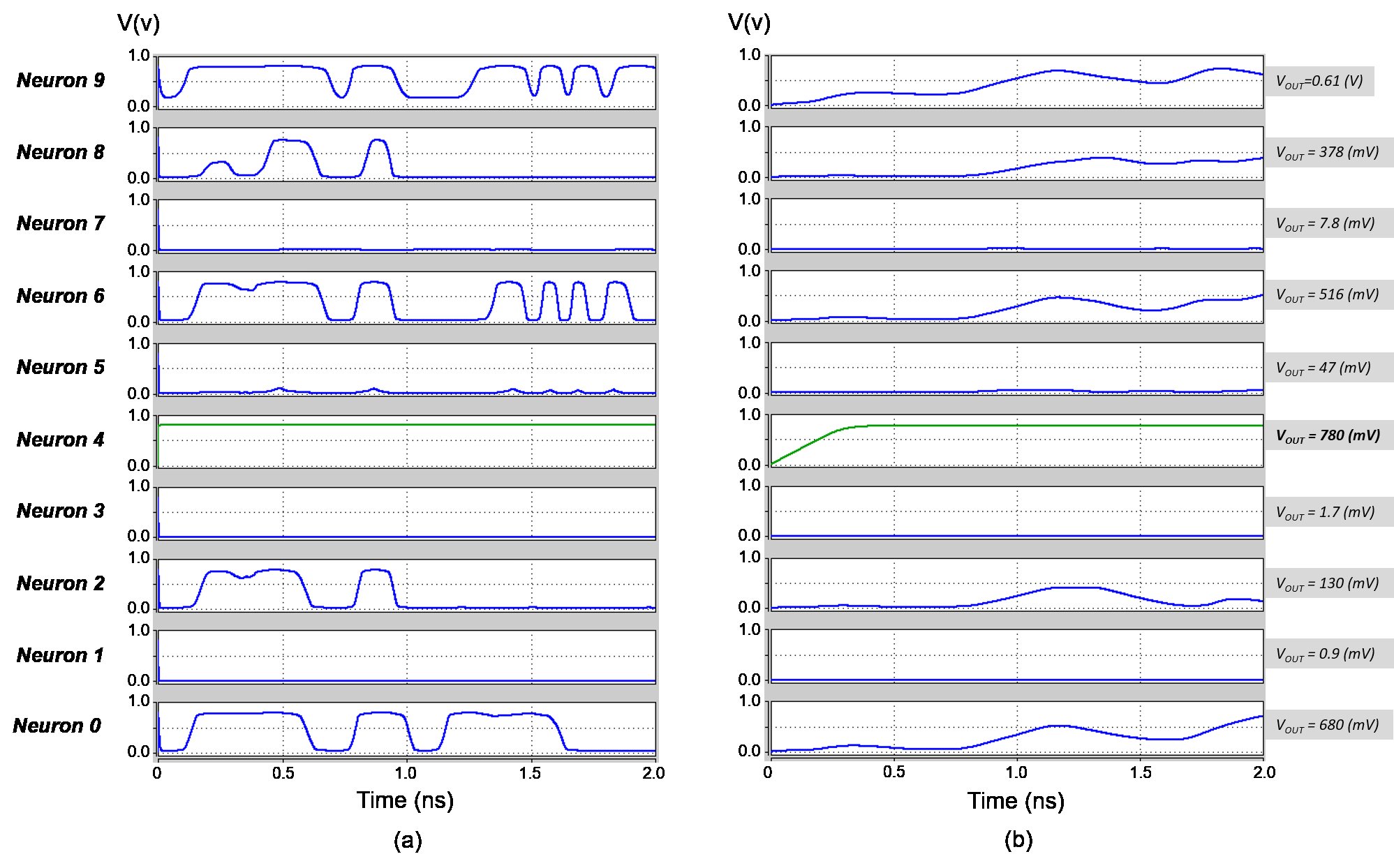}
\caption{Output of a $784\times200\times10$ DBN for a sample digit of ``4'' in the MNIST dataset: (a) Probabilistic output of the p-bit devices, (b) Output of the integrator circuit. The output voltage of the \textit{neuron-4}, which represents the digit ``4'' in the output classes, is greater than the other output voltages verifying a correct evaluation operation.}
\label{fig:digit4}
\end{figure}

\subsubsection{Power and Energy Consumption Analysis}
Figure~\ref{fig:pwr-eng}(a) depicts the power consumption of various DBN topologies while evaluating a single input image. As shown, a significant amount of power is consumed in the weighted connections, while less than 10\% of the total power is consumed in the neurons of an embedded MRAM-based p-bit approach. For instance, the total power consumption of a $784\times200\times10$ DBN is approximately equal to 86 mW, only 5.6 mW of which is dissipated in the activation functions. This is achieved by using the proposed power-efficient embedded MRAM-based neurons to implement the activation functions, as opposed to more elaborate floating-point circuits and pseudo-random number generators. Moreover, it is shown that the total power consumption depends primarily upon the aggregate number of neurons that are used in a network and not the number of layers. For instance, the power consumption of a $784\times500\times10$ DBN is greater than that of a $784\times200\times200\times10$ network, although the latter has higher number of hidden layers. However, the test operation delay is linearly proportional to the number of hidden layers which is determined by the signal propagation and computation progression. In particular, the RC integrator circuit shown in Figure~\ref{fig:rbm} is sampled every 2 ns, leading to an operating clock frequency of 500 MHz and a delay of 2 ns for each RBM. Thus, the $784\times200\times200\times10$ DBN mentioned above requires three clock cycles to complete the evaluation operation, while a $784\times500\times10$ DBN can produce its output in two clock cycles. Figure~\ref{fig:pwr-eng}(b) shows the energy consumption for various DBN topologies, which simultaneously includes the impact of number of nodes and hidden layers on power consumption and delay, respectively.

\begin{table}[]
\centering
\caption{PIN-Sim tunable parameters and their default values}
\label{tab:param}
\begin{tabular}{clc}
\hline
\textbf{Parameters}       & \textbf{Description}                                                & \textbf{Default Value} \\ \hline
$Topology$         & Defines the number of layers and nodes                            & $784\times200\times10$       \\
$TrainNum$                  & \# of training images                                                   & 3,000            \\
$R_{min}$       & Minimum resistance of the weighted connections                          & 1 $k\Omega$           \\
$\Delta R_W$ & Difference between min and max resistances of weighted connections & 400\%            \\
$Q$                         & Quantization factor                                                 & 8                \\
$R_0, R_1$                    & Resistances of the resistors in the differential amplifiers             & 1 $k\Omega$, 5 $k\Omega$    \\
$R_i, C_i$                    & Resistance and capacitance of the RC integrator circuits            & 100 $k\Omega$, 20 $fF$  \\ \hline
\end{tabular}
\end{table}

\subsubsection{PIN-Sim tunable parameters and their affect on DBN performance}
 Table~\ref{tab:param} lists the tunable parameters in the PIN-Sim framework, which can be adjusted based on the application requirements. The last column of the table shows the default values that are utilized herein for the MNIST digit recognition application. Figure~\ref{fig:digit4} shows the output voltages of the neurons in the last hidden layer of a $784\times200\times10$ DBN utilized for MNIST pattern recognition tasks, each of which represents an output class. The probabilistic outputs of the p-bit devices are shown in Figure~\ref{fig:digit4}(a), while Figure~\ref{fig:digit4}(b) exhibits the outputs of their corresponding integrator circuits. The outputs of the integrators are sampled after 2 ns, which is equal to the time constant of the integrator circuit. The output with the highest voltage amplitude represents the class to which the input image belongs. The results obtained exhibit a correct recognition operation for a sample input digit ``4'' within the MNIST dataset.

\begin{figure}
\centering
\includegraphics[scale=0.28]{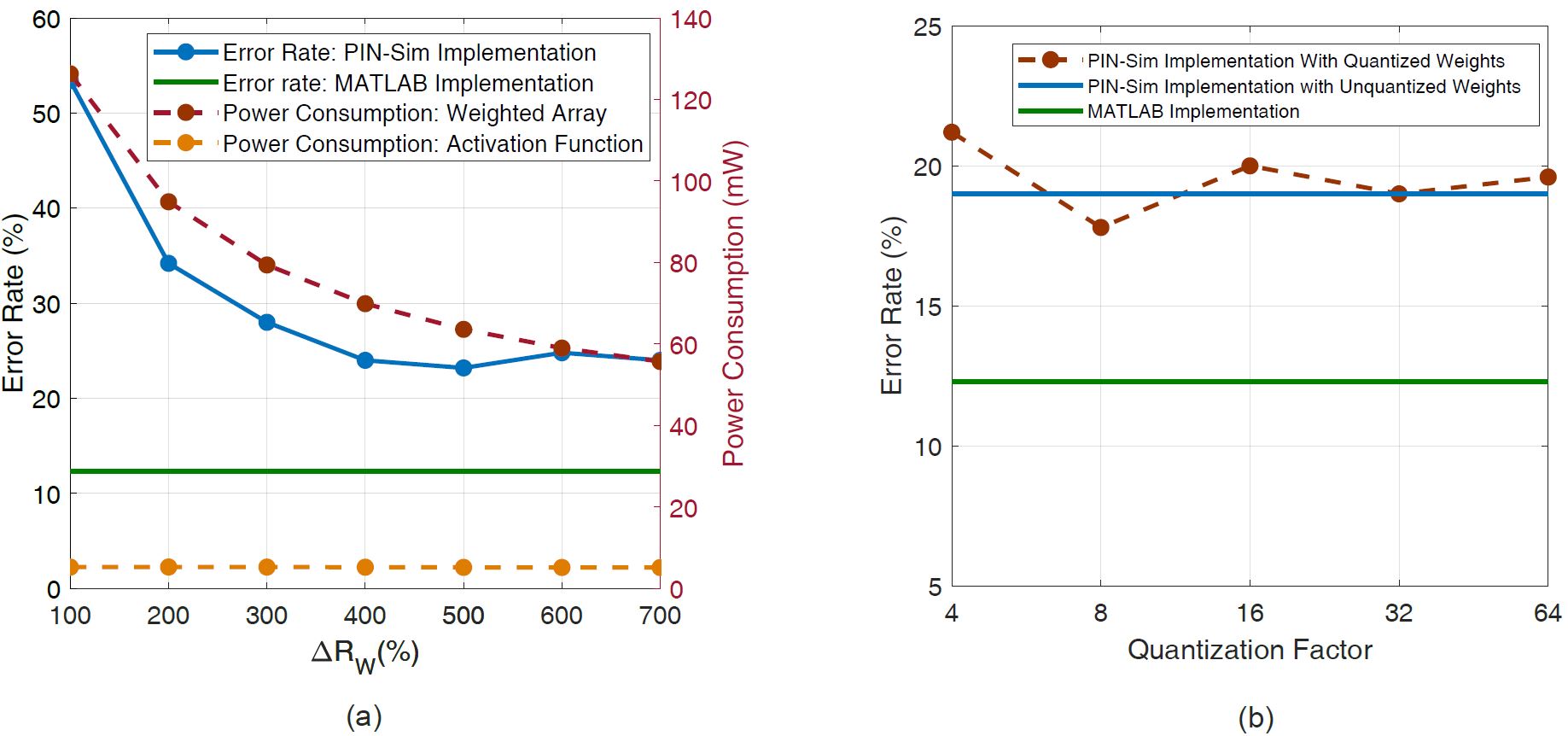}
\caption{(a) Error rate and power consumption versus $\Delta R_W$, and (b) error rate versus quantization factor (Q) for a $784\times200\times10$ DBN trained by 3,000 training images. The software implementation is technology-independent, in which the ideal sigmoid activation function and weight values are utilized in MATLAB to calculate the error rate. Thus, the changes in the tunable parameters used in the circuit-level SPICE implementation do not affect the measured error rates.}
\label{fig:tuneparam}
\end{figure}

Next, we will focus on the effect of some of the tunable parameters on the accuracy and power consumption of DBN architectures implemented by the proposed PIN-Sim framework. First, the effect of $\Delta R_W$ is investigated, which defines the possible resistance range of weights and biases as follows, $r_{max}=(1+\frac{\Delta R_W}{100})\times r_{min}$. The $r_{max}$ and $r_{min}$ parameters are utilized in the \textit{mapWEIGHT} module in the PIN-Sim tool to map the trained weights and biases to their corresponding resistance values according to Equations \ref{eqn:mapw1} and \ref{eqn:mapw2}, respectively. Figure \ref{fig:tuneparam}(a) shows the effect of $\Delta R_W$ on the recognition accuracy and power consumption of our default $784\times200\times10$ DBN implementation. As it can be seen in the figure, the error rate is reduced from 53\% to 24\% by increasing the $\Delta R_W$ from 100\% to 400\%, however a significant change in the error rate cannot be observed for $\Delta R_W$ values larger than 400\%. These results are particularly beneficial for magnetic tunnel junction (MTJ)-based weighted connections \cite{Sengupta2016hybrid,roy2018}, in which the difference between maximum and minimum resistance is defined by the tunneling magneto-resistance (TMR) effect. The results obtained show that a TMR of 400\% could be adequate to achieve the desired error rate. However, it is worth noting that this is quite application specific and can vary for different datasets. These results are worthy since the realization of higher TMR values would impose more complex fabrication processes \cite{parkin2004}, of which 700\% \cite{wang2009} have been demonstrated experimentally and others of 250\% \cite{wang2018} via current scalable means. Moreover, as it is shown in Figure \ref{fig:tuneparam}(a), increasing the $\Delta R_W$ results in reduced power dissipation in the weighted array, while the power dissipated in activation functions remains almost unchanged. The higher resistance range for the weighted connections increases the overall resistance of the weighted array. Therefore, since the input voltages remain unchanged the current flowing through the synapses will be decreased, which consequently reduces the power dissipated in the weighted array.

\begin{figure}
\centering
\includegraphics[scale=0.38]{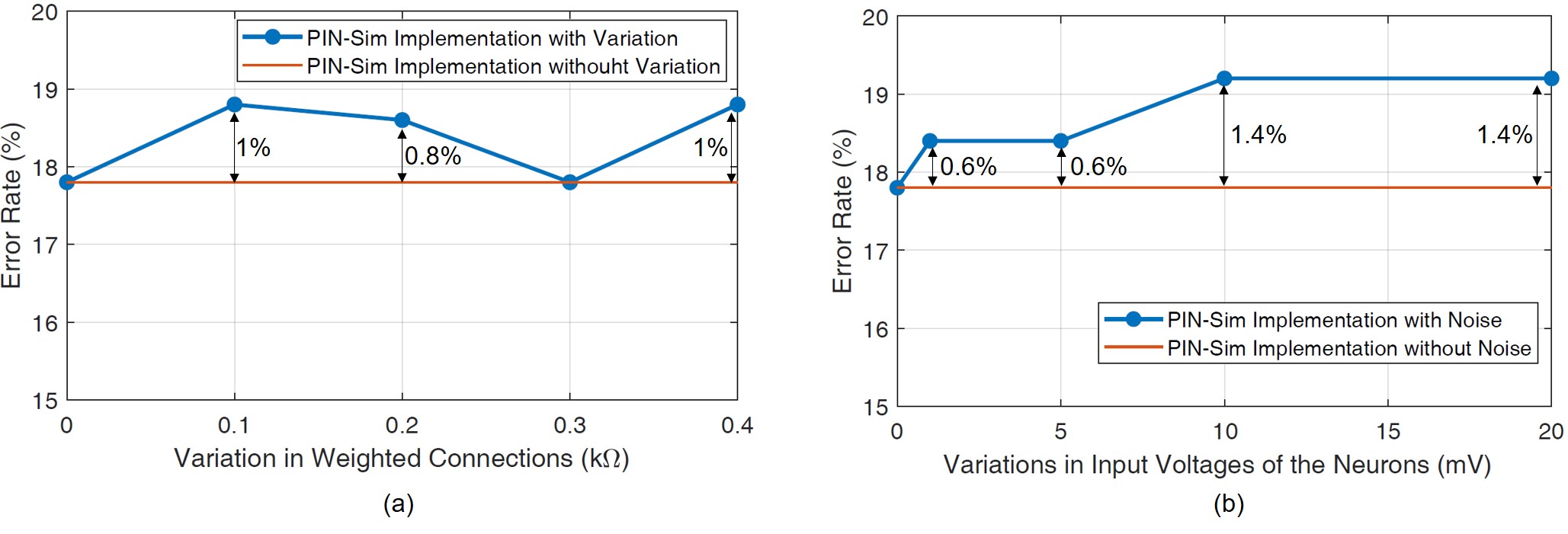}
\caption{(a) Error rate versus the variation in the resistance of weighted connections, and (b) error rate versus the variations in the input voltages of the neurons for a $784\times200\times10$ DBN trained by 3,000 training images.}
\label{fig:noisevar}
\end{figure}

In practice, providing an accurate and continuous range of weight resistances at nanoscale is not attainable due to the fabrication complexities and process variation. Therefore, a realistic circuit-level model of the resistive crossbar architecture should leverage quantized weights. Thus, leveraging PIN-Sim framework for design space exploration, we have assigned a quantization factor (Q) parameter, which can be tuned by the user based on the application requirements. Figure~\ref{fig:tuneparam}(b) shows the effect of weight discretization on the recognition accuracy of a $784\times200\times10$ DBN with $\Delta R_W$ of 400\% that is trained with 3,000 training samples. As shown, the error rate for the hardware implementation with $Q=4$, which means the weights are discretized into four equal intervals between $R_{min}$ and $R_{max}$, is increased to 21.2\% from the 19\% error rate that is achieved by the DBN with unquantized weights. As it is expected, this increase in the error rate is mainly caused by the information loss that occurs during the discretization. Moreover, implementations with larger $Q$ values result in error rates closer to that of the DBN with unquantized weights, which can also be expected since the discretization intervals are so small that the weight values are getting close to their unquantized values. However, an interesting phenomenon can be observed in the hardware implementation with $Q=8$, where the error rate of 17.8\% is realized which is lower than the error rate of the unquantized DBN. We have performed multiple tests to ensure that this is a repetitive behavior for the DBNs with $Q=8$, and in all of the cases the error rate obtained was lower than that of the DBN with unquantized weights. These results can be particularly interesting in the hardware-implementation, since for instance in our examined case there is a 0.5 $k\Omega$ gap between various weight resistances, considering the $R_{min}=1  k\Omega$ and $\Delta R_W=400\%$, which can provide some robustness against process variations without incurring a significant increase in the error rate. In particular, we have investigated the impacts of the variations in the input voltages of neurons, which can be induced by different noise sources, as well as variations in the resistance of the weighted connections on the recognition accuracy of the network. According to the results shown in Figure \ref{fig:noisevar} (a), a $784\times200\times10$ DBN trained by 3,000 images loses 1\% accuracy in presence of variations in weighted connections ranging from 0.1 $k\Omega$ to 0.4 $k\Omega$. Moreover, Figure \ref{fig:noisevar} (b) exhibits 1.4\% increase in the error rate for variations in the input voltages of neurons with a standard deviation of 20 mV.

\subsection{Discussion}
Some of the previous hardware implementations of DBNs are listed in Table~\ref{tab:compare}. The designs proposed in \cite{Kim2010,Ly2010} leverage FPGAs to achieve speedups of 25-145 compared to software implementations, however these approaches suffer from constrained clock frequencies and routing congestion, as well as major resource deficiencies due to the significant embedded memory utilization for both weighted connections and activation functions. In \cite{Yuan2017}, those authors have proposed optimization methods to reduce memory requirements for weights and biases, however implementing each activation function still requires dedicated piecewise linear approximator, random number generator (RNG), and comparator circuits which lead to increased area and energy consumption per neuron than the embedded MRAM-based approach herein. In \cite{Ardakani2017}, the low-complexity characteristics of stochastic CMOS-based arithmetic units are leveraged to implement RBM with reduced area and power consumption. However, the large number of linear feedback shift registers (LFSRs) that are required to generate the long input and weight bit-streams results in increased latencies that considerably limits the energy savings. 

\begin{table}[]
\centering
\caption{Various DBN hardware implementations with a focus on activation function structure.}
\label{tab:compare}
\small
\begin{tabular}{ccccc}
\hline
Design                               & Weighted Connection                                                             & Activation Function                                                                                               & \begin{tabular}[c]{@{}c@{}}Energy\\ per Neuron\end{tabular} & \begin{tabular}[c]{@{}c@{}}Normalized area\\ per neuron\end{tabular} \\ \hline
\cite{Kim2010}      & Embedded multipliers                                                            & CMOS-based LUTs                                                                                                   & N/A                                                         & N/A                                                                  \\ \hline
\cite{Ly2010}       & Embedded multipliers                                                            & \begin{tabular}[c]{@{}c@{}}-  2-kB BRAM\\ - Piecewise Linear Interpolator\\ - Random number Generator\end{tabular} & $\sim$10-100 nJ                                             & $\sim 3000 \times$                                                          \\ \hline
\cite{Yuan2017} & \begin{tabular}[c]{@{}c@{}}- Multiplier\\ - Adder tree \end{tabular} & \begin{tabular}[c]{@{}c@{}} - Piecewise Linear approximator\\ - Random number Generator \\- Comparator\end{tabular} & $\sim$10-100 nJ                                             & $\sim 2000 \times$  \\ \hline

\cite{Ardakani2017} & \begin{tabular}[c]{@{}c@{}}- LFSR\\ - bit-stream \\ - AND/OR gates\end{tabular} & \begin{tabular}[c]{@{}c@{}} -LFSR \\ - Bit-wise AND \\ - tree adder\\ - FSM-based \textit{tanh} unit\end{tabular} & $\sim$10-100 nJ                                             & $\sim 90 \times$  \\ \hline

\cite{SHERI2015}    & RRAM Memristor  & Off-chip   & N/A   & N/A  \\ \hline
\cite{Bojnordi2016} & RRAM  & \begin{tabular}[c]{@{}c@{}}- $64 \times 16$ LUTs\\ - Pseudo Random \\ Number Generator\\ - Comparator\end{tabular}            & $\sim$1-10 nJ                                               & $\sim 1250 \times$                                                          \\ \hline
\cite{Eryilmaz2016} & PCM & Off-chip    & N/A  & N/A  \\ \hline


Proposed Herein                      & Memristive Devices  & \begin{tabular}[c]{@{}c@{}}Embedded MRAM-based\\ Stocahstic Neuron \end{tabular} & \begin{tabular}[c]{@{}c@{}}Neuron: $\sim$1-10 fJ \\Integrator: $\sim$10-20 fJ\end{tabular}   & \begin{tabular}[c]{@{}c@{}}Neuron: $1\times$ \\Integrator: $\sim 3\times$\end{tabular} \\ \hline
\end{tabular}
\end{table}
On the other hand, emerging technologies such as resistive RAM (RRAM) and phase change memory (PCM) have been recently utilized within the crossbar arrays to implement matrix multiplication within RBMs \cite{Bojnordi2016,SHERI2015,Eryilmaz2016}. In particular, \cite{Bojnordi2016} has achieved $100\times$ and $10\times$ improvement in terms of operation speed and energy consumption, respectively, compared to single-threaded cores by using RRAM devices as weighted connections. In all of the above-mentioned designs, CMOS-based circuits such as multipliers and RNGs are utilized to realize the probabilistic behavior of activation functions. In \cite{zandRDBN}, authors have utilized low energy barrier spin-orbit torque (SOT) MTJs to implement the probabilistic sigmoidal activation function, which realizes significant area and energy reductions. However, the current-mode behavior of the SOT-MTJ devices imposes significant power consumption to the activation functions, while requiring weighted connections in $M\Omega$ resistances which can incur significant area overhead and fabrication complexity \cite{Sengupta2016hybrid,yuasa2004giant}. The work presented herein utilizes a voltage-driven embedded MRAM-based neuron with low energy barrier unstable nanomagnets, which leverages the intrinsic thermal noise to generate sigmoidal probabilistic activation functions required for RBMs within a power-efficient package. As listed in Table~\ref{tab:compare}, the proposed RBM implementation using embedded MRAM-based neuron can achieve approximately three orders of magnitude energy reduction compared to the previous energy-efficient CMOS-based implementations, while realizing at least ~90$\times$ device count reduction. However, as it was described in previous sections, the embedded MRAM based neuron requires an RC circuit to integrate its output voltage. The SPICE circuit simulation results exhibits an approximate average energy consumption of 10-20 fJ for the RC circuit as listed in Table~\ref{tab:compare}. Moreover, the area required to implement the RC circuit with 100 $K\Omega$ resistor and $20 fF$ capacitor is approximately three times larger than that of the MRAM-based neuron \cite{scott1998,stengel2006}. Thus, the proposed MRAM-based activation function can achieve approximately 20$\times$ and 300$\times$ area reduction compared to the CMOS-based stochastic neurons proposed in \cite{Ardakani2017} and \cite{Bojnordi2016}, respectively. The area of the MRAM-based neuron, which is utilized as the baseline for the area comparisons, is approximately equal to $32 \lambda \times 32 \lambda$, that is obtained by the layout design, in which $\lambda$ is a technology-dependent parameter. Herein, we have used the 14nm FinFET technology, which leads to the approximate area consumption of $0.05 \mu m^2$ per neuron. MRAM devices can be fabricated on top of the transistors, thus incurring near-zero area overhead.            
\vspace{-7.5pt}
\section{Conclusions}
Herein, it was shown that embedded MRAM-based neurons with thermally unstable superparamagnetic MTJs can realize a probabilistic output that can be modulated by an input voltage. The magnetization dynamics of the MRAM-based stochastic neuron is modeled by solving the LLG equations for a low energy barrier nanomagnet. The device-level simulations exhibited a desired sigmoidal relation between the input voltages and output probability of the neuron. Once the functionality of the proposed stochastic neuron was verified, we have developed an embedded MRAM-based RBM leveraging two resistive crossbar arrays with differential amplifiers to implement the matrix multiplication operation for both positive and negative weights. SPICE circuit simulations for a $2\times2$ weighted array validated the functionality of the proposed embedded MRAM-based RBM. 

To provide a circuit-level implementation of DBN, we have developed a PIN-Sim framework which is a transportable framework for rapid, automated, and accurate design space exploration of hybrid CMOS and post-CMOS neuromorphic circuits. PIN-Sim is composed of five main modules to train the network, map the trained weights to their corresponding resistances, create the SPICE model of the RBMs, and measure the accuracy and energy consumption. MNIST dataset is utilized to investigate the accuracy and energy tradeoffs for seven distinct DBN topologies implemented by the developed PIN-Sim framework. The simulation results showed that at least two hidden layers are required to achieve suitable error rates. In particular, a $784\times200\times10$ DBN can realize 5\% error rate while consuming less than 500 pJ energy. The error rates could be decreased to 2.5\% by using a $784 \times 500 \times 500 \times 500 \times 10$ DBN topologies trained by 10,000 input training samples at the cost of $\sim 10\times$ higher energy consumption and significantly larger area overheads. Moreover, PIN-Sim can be used to optimize network topologies based on different application requirements for energy versus accuracy tradeoffs.

Next, we have focused on the effect of various hardware-level parameters that can be adjusted in the PIN-Sim tool on the performance of the network. One particular parameter which is specifically important for MTJ and RRAM based crossbar architectures is the difference between the largest and smallest possible resistance values in a weighted connection ($\Delta R_W$). It was shown that at least a $\Delta R_W$ of 400\% is required to realize suitable error rates, however it is worth noting that increasing the $\Delta R_W$ to values larger than 400\% does not lead to a significant reduction in error rate. Therefore, some fabrication complexities for increasing the $\Delta R_W$ in MTJ-based weighted connections can be avoided. Moreover, to realize a realistic hardware implementation we have studied the effect of weight quantization on the accuracy of our network. It was shown that a quantization factor of eight, which provides eight different resistive levels in each weighted connection, can lead to even lower error rates compared to a network with unquantized weights. This also shows the robustness of our proposed circuit-level DBN implementation to minor variations in the resistance of the weighted connections, which is inevitable during the fabrication process. Finally, the comparison results exhibited that the embedded MRAM-based neuron can contribute to several orders of magnitude energy reduction, and reduce the area requirement by 20-fold, with respect to recent energy-optimized designs. Although this is a simulation-based result, hardware realization may endure significant process variation and impacts of sneak currents in large crossbar arrays. While on-chip training can help to mitigate these somewhat, alternate approaches using binarized weights are options explored in other works with varying results \cite{Courbariaux2015}. To address these further, the development of the PIN-Sim framework provides several possibilities for future work, including: (1) leveraging optimization techniques to reduce the performance gap between the ideal implementation of the DBN using simulation tools such as MATLAB, and the realistic circuit-level implementation of DBN using PIN-Sim framework, (2) training DBNs with binary weights which can be implemented by MTJs or RRAMs, (3) implementing convolutional DBNs using PIN-Sim for more complex pattern recognition applications.\\

\vspace{-10pt}

\begin{acks}
 This work was supported in part by the Center for Probabilistic Spin Logic for Low-Energy Boolean and Non-Boolean Computing  (CAPSL), one of the Nanoelectronic Computing Research (nCORE) Centers as task 2759.006, a Semiconductor Research Corporation (SRC) program sponsored by the NSF through CCF 1739635.
\end{acks}
\vspace{-10pt}
\bibliographystyle{ACM-Reference-Format}
\bibliography{sample-bibliography}


\begin{thebibliography}{55}


\ifx \showCODEN    \undefined \def \showCODEN     #1{\unskip}     \fi
\ifx \showDOI      \undefined \def \showDOI       #1{#1}\fi
\ifx \showISBNx    \undefined \def \showISBNx     #1{\unskip}     \fi
\ifx \showISBNxiii \undefined \def \showISBNxiii  #1{\unskip}     \fi
\ifx \showISSN     \undefined \def \showISSN      #1{\unskip}     \fi
\ifx \showLCCN     \undefined \def \showLCCN      #1{\unskip}     \fi
\ifx \shownote     \undefined \def \shownote      #1{#1}          \fi
\ifx \showarticletitle \undefined \def \showarticletitle #1{#1}   \fi
\ifx \showURL      \undefined \def \showURL       {\relax}        \fi
\providecommand\bibfield[2]{#2}
\providecommand\bibinfo[2]{#2}
\providecommand\natexlab[1]{#1}
\providecommand\showeprint[2][]{arXiv:#2}

\bibitem[\protect\citeauthoryear{??}{pre}{[n. d.]}]%
        {predictive_tech}
 \bibinfo{year}{[n. d.]}\natexlab{}.
\newblock \bibinfo{title}{Predictive {Technology} {Model} ({PTM})
  (http://ptm.asu.edu/)}.
\newblock
\newblock


\bibitem[\protect\citeauthoryear{Ackley, Hinton, and Sejnowski}{Ackley
  et~al\mbox{.}}{1985}]%
        {ackley1985}
\bibfield{author}{\bibinfo{person}{David~H Ackley}, \bibinfo{person}{Geoffrey~E
  Hinton}, {and} \bibinfo{person}{Terrence~J Sejnowski}.}
  \bibinfo{year}{1985}\natexlab{}.
\newblock \showarticletitle{A learning algorithm for Boltzmann machines}.
\newblock \bibinfo{journal}{\emph{Cognitive science}} \bibinfo{volume}{9},
  \bibinfo{number}{1} (\bibinfo{year}{1985}), \bibinfo{pages}{147--169}.
\newblock


\bibitem[\protect\citeauthoryear{Ardakani, Leduc-Primeau, Onizawa, Hanyu, and
  Gross}{Ardakani et~al\mbox{.}}{2017}]%
        {Ardakani2017}
\bibfield{author}{\bibinfo{person}{A. Ardakani}, \bibinfo{person}{F.
  Leduc-Primeau}, \bibinfo{person}{N. Onizawa}, \bibinfo{person}{T. Hanyu},
  {and} \bibinfo{person}{W.~J. Gross}.} \bibinfo{year}{2017}\natexlab{}.
\newblock \showarticletitle{VLSI Implementation of Deep Neural Network Using
  Integral Stochastic Computing}.
\newblock \bibinfo{journal}{\emph{IEEE Transactions on Very Large Scale
  Integration (VLSI) Systems}} \bibinfo{volume}{25}, \bibinfo{number}{10}
  (\bibinfo{year}{2017}).
\newblock
\showISSN{1063-8210}


\bibitem[\protect\citeauthoryear{Bapna and Majetich}{Bapna and
  Majetich}{2017}]%
        {bapna2017current}
\bibfield{author}{\bibinfo{person}{Mukund Bapna} {and} \bibinfo{person}{Sara~A
  Majetich}.} \bibinfo{year}{2017}\natexlab{}.
\newblock \showarticletitle{Current control of time-averaged magnetization in
  superparamagnetic tunnel junctions}.
\newblock \bibinfo{journal}{\emph{Applied Physics Letters}}
  \bibinfo{volume}{111}, \bibinfo{number}{24} (\bibinfo{year}{2017}),
  \bibinfo{pages}{243107}.
\newblock


\bibitem[\protect\citeauthoryear{Basheer and Hajmeer}{Basheer and
  Hajmeer}{2000}]%
        {BASHEER2000}
\bibfield{author}{\bibinfo{person}{I.A Basheer} {and} \bibinfo{person}{M
  Hajmeer}.} \bibinfo{year}{2000}\natexlab{}.
\newblock \showarticletitle{Artificial neural networks: fundamentals,
  computing, design, and application}.
\newblock \bibinfo{journal}{\emph{Journal of Microbiological Methods}}
  \bibinfo{volume}{43}, \bibinfo{number}{1} (\bibinfo{year}{2000}),
  \bibinfo{pages}{3 -- 31}.
\newblock
\showISSN{0167-7012}
\urldef\tempurl%
\url{https://doi.org/10.1016/S0167-7012(00)00201-3}
\showDOI{\tempurl}
\newblock
\shownote{Neural Computting in Micrbiology.}


\bibitem[\protect\citeauthoryear{Behin-Aein, Diep, and Datta}{Behin-Aein
  et~al\mbox{.}}{2016}]%
        {behin2016}
\bibfield{author}{\bibinfo{person}{Behtash Behin-Aein}, \bibinfo{person}{Vinh
  Diep}, {and} \bibinfo{person}{Supriyo Datta}.}
  \bibinfo{year}{2016}\natexlab{}.
\newblock \showarticletitle{A building block for hardware belief networks}.
\newblock \bibinfo{journal}{\emph{Scientific reports}}  \bibinfo{volume}{6}
  (\bibinfo{year}{2016}).
\newblock


\bibitem[\protect\citeauthoryear{Bhatti, Sbiaa, Hirohata, Ohno, Fukami, and
  Piramanayagam}{Bhatti et~al\mbox{.}}{2017}]%
        {bhatti2017spintronics}
\bibfield{author}{\bibinfo{person}{Sabpreet Bhatti}, \bibinfo{person}{Rachid
  Sbiaa}, \bibinfo{person}{Atsufumi Hirohata}, \bibinfo{person}{Hideo Ohno},
  \bibinfo{person}{Shunsuke Fukami}, {and} \bibinfo{person}{SN Piramanayagam}.}
  \bibinfo{year}{2017}\natexlab{}.
\newblock \showarticletitle{Spintronics based random access memory: a review}.
\newblock \bibinfo{journal}{\emph{Materials Today}} (\bibinfo{year}{2017}).
\newblock


\bibitem[\protect\citeauthoryear{Bishop, Bishop, et~al\mbox{.}}{Bishop
  et~al\mbox{.}}{1995}]%
        {bishop1995}
\bibfield{author}{\bibinfo{person}{Chris Bishop},
  \bibinfo{person}{Christopher~M Bishop}, {et~al\mbox{.}}}
  \bibinfo{year}{1995}\natexlab{}.
\newblock \bibinfo{booktitle}{\emph{Neural networks for pattern recognition}}.
\newblock \bibinfo{publisher}{Oxford university press}.
\newblock


\bibitem[\protect\citeauthoryear{Bojnordi and Ipek}{Bojnordi and Ipek}{2016}]%
        {Bojnordi2016}
\bibfield{author}{\bibinfo{person}{M.~N. Bojnordi} {and} \bibinfo{person}{E.
  Ipek}.} \bibinfo{year}{2016}\natexlab{}.
\newblock \showarticletitle{Memristive Boltzmann machine: A hardware
  accelerator for combinatorial optimization and deep learning}. In
  \bibinfo{booktitle}{\emph{2016 IEEE International Symposium on High
  Performance Computer Architecture (HPCA)}}.
\newblock


\bibitem[\protect\citeauthoryear{Buesing, Bill, Nessler, and Maass}{Buesing
  et~al\mbox{.}}{2011}]%
        {buesing2011}
\bibfield{author}{\bibinfo{person}{Lars Buesing}, \bibinfo{person}{Johannes
  Bill}, \bibinfo{person}{Bernhard Nessler}, {and} \bibinfo{person}{Wolfgang
  Maass}.} \bibinfo{year}{2011}\natexlab{}.
\newblock \showarticletitle{Neural dynamics as sampling: a model for stochastic
  computation in recurrent networks of spiking neurons}.
\newblock \bibinfo{journal}{\emph{PLoS computational biology}}
  \bibinfo{volume}{7}, \bibinfo{number}{11} (\bibinfo{year}{2011}),
  \bibinfo{pages}{e1002211}.
\newblock


\bibitem[\protect\citeauthoryear{Camsari, Faria, Sutton, and Datta}{Camsari
  et~al\mbox{.}}{2017a}]%
        {camsari2017stochastic}
\bibfield{author}{\bibinfo{person}{Kerem~Yunus Camsari},
  \bibinfo{person}{Rafatul Faria}, \bibinfo{person}{Brian~M Sutton}, {and}
  \bibinfo{person}{Supriyo Datta}.} \bibinfo{year}{2017}\natexlab{a}.
\newblock \showarticletitle{Stochastic p-bits for invertible logic}.
\newblock \bibinfo{journal}{\emph{Physical Review X}} \bibinfo{volume}{7},
  \bibinfo{number}{3} (\bibinfo{year}{2017}), \bibinfo{pages}{031014}.
\newblock


\bibitem[\protect\citeauthoryear{Camsari, Ganguly, and Datta}{Camsari
  et~al\mbox{.}}{2015}]%
        {camsari2015modular}
\bibfield{author}{\bibinfo{person}{Kerem~Yunus Camsari},
  \bibinfo{person}{Samiran Ganguly}, {and} \bibinfo{person}{Supriyo Datta}.}
  \bibinfo{year}{2015}\natexlab{}.
\newblock \showarticletitle{Modular approach to spintronics}.
\newblock \bibinfo{journal}{\emph{Scientific reports}}  \bibinfo{volume}{5}
  (\bibinfo{year}{2015}), \bibinfo{pages}{10571}.
\newblock


\bibitem[\protect\citeauthoryear{Camsari, Salahuddin, and Datta}{Camsari
  et~al\mbox{.}}{2017b}]%
        {camsari2017implementing}
\bibfield{author}{\bibinfo{person}{Kerem~Yunus Camsari},
  \bibinfo{person}{Sayeef Salahuddin}, {and} \bibinfo{person}{Supriyo Datta}.}
  \bibinfo{year}{2017}\natexlab{b}.
\newblock \showarticletitle{Implementing p-bits with embedded mtj}.
\newblock \bibinfo{journal}{\emph{IEEE Electron Device Letters}}
  \bibinfo{volume}{38}, \bibinfo{number}{12} (\bibinfo{year}{2017}),
  \bibinfo{pages}{1767--1770}.
\newblock


\bibitem[\protect\citeauthoryear{Carreira-Perpinan and
  Hinton}{Carreira-Perpinan and Hinton}{2005}]%
        {carreira2005}
\bibfield{author}{\bibinfo{person}{Miguel~A Carreira-Perpinan} {and}
  \bibinfo{person}{Geoffrey~E Hinton}.} \bibinfo{year}{2005}\natexlab{}.
\newblock \showarticletitle{On contrastive divergence learning.}. In
  \bibinfo{booktitle}{\emph{Aistats}}, Vol.~\bibinfo{volume}{10}.
  \bibinfo{pages}{33--40}.
\newblock


\bibitem[\protect\citeauthoryear{Choi, Lv, Kim, Deshpande, Kang, Wang, and
  Kim}{Choi et~al\mbox{.}}{2014}]%
        {choi2014magnetic}
\bibfield{author}{\bibinfo{person}{Won~Ho Choi}, \bibinfo{person}{Yang Lv},
  \bibinfo{person}{Jongyeon Kim}, \bibinfo{person}{Abhishek Deshpande},
  \bibinfo{person}{Gyuseong Kang}, \bibinfo{person}{Jian-Ping Wang}, {and}
  \bibinfo{person}{Chris~H Kim}.} \bibinfo{year}{2014}\natexlab{}.
\newblock \showarticletitle{A magnetic tunnel junction based true random number
  generator with conditional perturb and real-time output probability
  tracking}. In \bibinfo{booktitle}{\emph{Electron Devices Meeting (IEDM), 2014
  IEEE International}}. IEEE, \bibinfo{pages}{12--5}.
\newblock


\bibitem[\protect\citeauthoryear{Churchland and Sejnowski}{Churchland and
  Sejnowski}{2016}]%
        {churchland2016}
\bibfield{author}{\bibinfo{person}{Patricia~S Churchland} {and}
  \bibinfo{person}{Terrence~J Sejnowski}.} \bibinfo{year}{2016}\natexlab{}.
\newblock \bibinfo{booktitle}{\emph{The computational brain}}.
\newblock \bibinfo{publisher}{MIT press}.
\newblock


\bibitem[\protect\citeauthoryear{Courbariaux, Bengio, and David}{Courbariaux
  et~al\mbox{.}}{2015}]%
        {Courbariaux2015}
\bibfield{author}{\bibinfo{person}{Matthieu Courbariaux},
  \bibinfo{person}{Yoshua Bengio}, {and} \bibinfo{person}{Jean-Pierre David}.}
  \bibinfo{year}{2015}\natexlab{}.
\newblock \showarticletitle{BinaryConnect: Training Deep Neural Networks with
  binary weights during propagations}.
\newblock In \bibinfo{booktitle}{\emph{Advances in Neural Information
  Processing Systems 28}}, \bibfield{editor}{\bibinfo{person}{C.~Cortes},
  \bibinfo{person}{N.~D. Lawrence}, \bibinfo{person}{D.~D. Lee},
  \bibinfo{person}{M.~Sugiyama}, {and} \bibinfo{person}{R.~Garnett}} (Eds.).
  \bibinfo{publisher}{Curran Associates, Inc.}, \bibinfo{pages}{3123--3131}.
\newblock
\urldef\tempurl%
\url{http://papers.nips.cc/paper/5647-binaryconnect-training-deep-neural-networks-with-binary-weights-during-propagations.pdf}
\showURL{%
\tempurl}


\bibitem[\protect\citeauthoryear{Cowburn, Koltsov, Adeyeye, Welland, and
  Tricker}{Cowburn et~al\mbox{.}}{1999}]%
        {Cowburn1999}
\bibfield{author}{\bibinfo{person}{R.~P. Cowburn}, \bibinfo{person}{D.~K.
  Koltsov}, \bibinfo{person}{A.~O. Adeyeye}, \bibinfo{person}{M.~E. Welland},
  {and} \bibinfo{person}{D.~M. Tricker}.} \bibinfo{year}{1999}\natexlab{}.
\newblock \showarticletitle{Single-Domain Circular Nanomagnets}.
\newblock \bibinfo{journal}{\emph{Phys. Rev. Lett.}}  \bibinfo{volume}{83}
  (\bibinfo{date}{Aug} \bibinfo{year}{1999}), \bibinfo{pages}{1042--1045}.
\newblock
Issue 5.
\urldef\tempurl%
\url{https://doi.org/10.1103/PhysRevLett.83.1042}
\showDOI{\tempurl}


\bibitem[\protect\citeauthoryear{Debashis, Faria, Camsari, Appenzeller, Datta,
  and Chen}{Debashis et~al\mbox{.}}{2016}]%
        {debashis2016experimental}
\bibfield{author}{\bibinfo{person}{Punyashloka Debashis},
  \bibinfo{person}{Rafatul Faria}, \bibinfo{person}{Kerem~Y Camsari},
  \bibinfo{person}{Joerg Appenzeller}, \bibinfo{person}{Supriyo Datta}, {and}
  \bibinfo{person}{Zhihong Chen}.} \bibinfo{year}{2016}\natexlab{}.
\newblock \showarticletitle{Experimental demonstration of nanomagnet networks
  as hardware for ising computing}. In \bibinfo{booktitle}{\emph{Electron
  Devices Meeting (IEDM), 2016 IEEE International}}. IEEE,
  \bibinfo{pages}{34--3}.
\newblock


\bibitem[\protect\citeauthoryear{Debashis, Faria, Camsari, and Chen}{Debashis
  et~al\mbox{.}}{2018}]%
        {Debashis2018}
\bibfield{author}{\bibinfo{person}{P. Debashis}, \bibinfo{person}{R. Faria},
  \bibinfo{person}{K.~Y. Camsari}, {and} \bibinfo{person}{Z. Chen}.}
  \bibinfo{year}{2018}\natexlab{}.
\newblock \showarticletitle{Design of Stochastic Nanomagnets for Probabilistic
  Spin Logic}.
\newblock \bibinfo{journal}{\emph{IEEE Magnetics Letters}}  \bibinfo{volume}{9}
  (\bibinfo{year}{2018}), \bibinfo{pages}{1--5}.
\newblock
\showISSN{1949-307X}
\urldef\tempurl%
\url{https://doi.org/10.1109/LMAG.2018.2860547}
\showDOI{\tempurl}


\bibitem[\protect\citeauthoryear{Eryilmaz, Neftci, Joshi, Kim, BrightSky, Lung,
  Lam, Cauwenberghs, and Wong}{Eryilmaz et~al\mbox{.}}{2016}]%
        {Eryilmaz2016}
\bibfield{author}{\bibinfo{person}{S.~B. Eryilmaz}, \bibinfo{person}{E.
  Neftci}, \bibinfo{person}{S. Joshi}, \bibinfo{person}{S. Kim},
  \bibinfo{person}{M. BrightSky}, \bibinfo{person}{H.~L. Lung},
  \bibinfo{person}{C. Lam}, \bibinfo{person}{G. Cauwenberghs}, {and}
  \bibinfo{person}{H.~S.~P. Wong}.} \bibinfo{year}{2016}\natexlab{}.
\newblock \showarticletitle{Training a Probabilistic Graphical Model With
  Resistive Switching Electronic Synapses}.
\newblock \bibinfo{journal}{\emph{IEEE Transactions on Electron Devices}}
  \bibinfo{volume}{63}, \bibinfo{number}{12} (\bibinfo{date}{Dec}
  \bibinfo{year}{2016}), \bibinfo{pages}{5004--5011}.
\newblock
\showISSN{0018-9383}


\bibitem[\protect\citeauthoryear{Fukushima, Seki, Yakushiji, Kubota, Imamura,
  Yuasa, and Ando}{Fukushima et~al\mbox{.}}{2014}]%
        {fukushima2014spin}
\bibfield{author}{\bibinfo{person}{Akio Fukushima}, \bibinfo{person}{Takayuki
  Seki}, \bibinfo{person}{Kay Yakushiji}, \bibinfo{person}{Hitoshi Kubota},
  \bibinfo{person}{Hiroshi Imamura}, \bibinfo{person}{Shinji Yuasa}, {and}
  \bibinfo{person}{Koji Ando}.} \bibinfo{year}{2014}\natexlab{}.
\newblock \showarticletitle{Spin dice: A scalable truly random number generator
  based on spintronics}.
\newblock \bibinfo{journal}{\emph{Applied Physics Express}}
  \bibinfo{volume}{7}, \bibinfo{number}{8} (\bibinfo{year}{2014}),
  \bibinfo{pages}{083001}.
\newblock


\bibitem[\protect\citeauthoryear{Hecht-Nielsen}{Hecht-Nielsen}{1992}]%
        {hecht1992}
\bibfield{author}{\bibinfo{person}{Robert Hecht-Nielsen}.}
  \bibinfo{year}{1992}\natexlab{}.
\newblock \showarticletitle{Theory of the backpropagation neural network}.
\newblock In \bibinfo{booktitle}{\emph{Neural networks for perception}}.
  \bibinfo{publisher}{Elsevier}, \bibinfo{pages}{65--93}.
\newblock


\bibitem[\protect\citeauthoryear{Hinton, Osindero, and Teh}{Hinton
  et~al\mbox{.}}{2006}]%
        {hinton2006}
\bibfield{author}{\bibinfo{person}{Geoffrey~E Hinton}, \bibinfo{person}{Simon
  Osindero}, {and} \bibinfo{person}{Yee-Whye Teh}.}
  \bibinfo{year}{2006}\natexlab{}.
\newblock \showarticletitle{A fast learning algorithm for deep belief nets}.
\newblock \bibinfo{journal}{\emph{Neural computation}} \bibinfo{volume}{18},
  \bibinfo{number}{7} (\bibinfo{year}{2006}), \bibinfo{pages}{1527--1554}.
\newblock


\bibitem[\protect\citeauthoryear{Hinton, Sejnowski, and Ackley}{Hinton
  et~al\mbox{.}}{1984}]%
        {hinton1984boltzmann}
\bibfield{author}{\bibinfo{person}{Geoffrey~E Hinton},
  \bibinfo{person}{Terrence~J Sejnowski}, {and} \bibinfo{person}{David~H
  Ackley}.} \bibinfo{year}{1984}\natexlab{}.
\newblock \bibinfo{booktitle}{\emph{Boltzmann machines: Constraint satisfaction
  networks that learn}}.
\newblock \bibinfo{publisher}{Carnegie-Mellon University, Department of
  Computer Science Pittsburgh, PA}.
\newblock


\bibitem[\protect\citeauthoryear{Hu, Li, Wu, and Rose}{Hu
  et~al\mbox{.}}{2012}]%
        {Hu2012}
\bibfield{author}{\bibinfo{person}{Miao Hu}, \bibinfo{person}{Hai Li},
  \bibinfo{person}{Qing Wu}, {and} \bibinfo{person}{Garrett~S. Rose}.}
  \bibinfo{year}{2012}\natexlab{}.
\newblock \showarticletitle{Hardware Realization of BSB Recall Function Using
  Memristor Crossbar Arrays}. In \bibinfo{booktitle}{\emph{Proceedings of the
  49th Annual Design Automation Conference}} \emph{(\bibinfo{series}{DAC
  '12})}. \bibinfo{publisher}{ACM}, \bibinfo{address}{New York, NY, USA},
  \bibinfo{pages}{498--503}.
\newblock
\showISBNx{978-1-4503-1199-1}
\urldef\tempurl%
\url{https://doi.org/10.1145/2228360.2228448}
\showDOI{\tempurl}


\bibitem[\protect\citeauthoryear{Kim, McMahon, and Olukotun}{Kim
  et~al\mbox{.}}{2010}]%
        {Kim2010}
\bibfield{author}{\bibinfo{person}{S.~K. Kim}, \bibinfo{person}{P.~L. McMahon},
  {and} \bibinfo{person}{K. Olukotun}.} \bibinfo{year}{2010}\natexlab{}.
\newblock \showarticletitle{A Large-Scale Architecture for Restricted Boltzmann
  Machines}. In \bibinfo{booktitle}{\emph{2010 18th IEEE Annual International
  Symposium on Field-Programmable Custom Computing Machines}}.
  \bibinfo{pages}{201--208}.
\newblock


\bibitem[\protect\citeauthoryear{LeCun, Bengio, and Hinton}{LeCun
  et~al\mbox{.}}{2015}]%
        {lecun2015deep}
\bibfield{author}{\bibinfo{person}{Yann LeCun}, \bibinfo{person}{Yoshua
  Bengio}, {and} \bibinfo{person}{Geoffrey Hinton}.}
  \bibinfo{year}{2015}\natexlab{}.
\newblock \showarticletitle{Deep learning}.
\newblock \bibinfo{journal}{\emph{nature}} \bibinfo{volume}{521},
  \bibinfo{number}{7553} (\bibinfo{year}{2015}), \bibinfo{pages}{436}.
\newblock


\bibitem[\protect\citeauthoryear{Lecun, Bottou, Bengio, and Haffner}{Lecun
  et~al\mbox{.}}{1998}]%
        {Lecun1998}
\bibfield{author}{\bibinfo{person}{Y. Lecun}, \bibinfo{person}{L. Bottou},
  \bibinfo{person}{Y. Bengio}, {and} \bibinfo{person}{P. Haffner}.}
  \bibinfo{year}{1998}\natexlab{}.
\newblock \showarticletitle{Gradient-based learning applied to document
  recognition}.
\newblock \bibinfo{journal}{\emph{Proc. IEEE}} \bibinfo{volume}{86},
  \bibinfo{number}{11} (\bibinfo{date}{Nov} \bibinfo{year}{1998}),
  \bibinfo{pages}{2278--2324}.
\newblock
\showISSN{0018-9219}
\urldef\tempurl%
\url{https://doi.org/10.1109/5.726791}
\showDOI{\tempurl}


\bibitem[\protect\citeauthoryear{Lin, Kang, Wang, Lee, Zhu, Chen, Li, Hsu, Kao,
  Liu, et~al\mbox{.}}{Lin et~al\mbox{.}}{2009}]%
        {lin2009}
\bibfield{author}{\bibinfo{person}{CJ Lin}, \bibinfo{person}{SH Kang},
  \bibinfo{person}{YJ Wang}, \bibinfo{person}{K Lee}, \bibinfo{person}{X Zhu},
  \bibinfo{person}{WC Chen}, \bibinfo{person}{X Li}, \bibinfo{person}{WN Hsu},
  \bibinfo{person}{YC Kao}, \bibinfo{person}{MT Liu}, {et~al\mbox{.}}}
  \bibinfo{year}{2009}\natexlab{}.
\newblock \showarticletitle{45nm low power CMOS logic compatible embedded STT
  MRAM utilizing a reverse-connection 1T/1MTJ cell}. In
  \bibinfo{booktitle}{\emph{Electron Devices Meeting (IEDM), 2009 IEEE
  International}}. IEEE, \bibinfo{pages}{1--4}.
\newblock


\bibitem[\protect\citeauthoryear{Liyanagedera, Sengupta, Jaiswal, and
  Roy}{Liyanagedera et~al\mbox{.}}{2017}]%
        {liyanagedera2017magnetic}
\bibfield{author}{\bibinfo{person}{Chamika~M Liyanagedera},
  \bibinfo{person}{Abhronil Sengupta}, \bibinfo{person}{Akhilesh Jaiswal},
  {and} \bibinfo{person}{Kaushik Roy}.} \bibinfo{year}{2017}\natexlab{}.
\newblock \showarticletitle{Magnetic tunnel junction enabled stochastic spiking
  neural networks: From non-telegraphic to telegraphic switching regime}.
\newblock \bibinfo{journal}{\emph{arXiv preprint arXiv:1709.09247}}
  (\bibinfo{year}{2017}).
\newblock


\bibitem[\protect\citeauthoryear{Locatelli, Mizrahi, Accioly, Matsumoto,
  Fukushima, Kubota, Yuasa, Cros, Pereira, Querlioz, et~al\mbox{.}}{Locatelli
  et~al\mbox{.}}{2014}]%
        {locatelli2014noise}
\bibfield{author}{\bibinfo{person}{Nicolas Locatelli}, \bibinfo{person}{Alice
  Mizrahi}, \bibinfo{person}{A Accioly}, \bibinfo{person}{Rie Matsumoto},
  \bibinfo{person}{Akio Fukushima}, \bibinfo{person}{Hitoshi Kubota},
  \bibinfo{person}{Shinji Yuasa}, \bibinfo{person}{Vincent Cros},
  \bibinfo{person}{Luis~Gustavo Pereira}, \bibinfo{person}{Damien Querlioz},
  {et~al\mbox{.}}} \bibinfo{year}{2014}\natexlab{}.
\newblock \showarticletitle{Noise-enhanced synchronization of stochastic
  magnetic oscillators}.
\newblock \bibinfo{journal}{\emph{Physical Review Applied}}
  \bibinfo{volume}{2}, \bibinfo{number}{3} (\bibinfo{year}{2014}),
  \bibinfo{pages}{034009}.
\newblock


\bibitem[\protect\citeauthoryear{Ly and Chow}{Ly and Chow}{2010}]%
        {Ly2010}
\bibfield{author}{\bibinfo{person}{D.~L. Ly} {and} \bibinfo{person}{P. Chow}.}
  \bibinfo{year}{2010}\natexlab{}.
\newblock \showarticletitle{High-Performance Reconfigurable Hardware
  Architecture for Restricted Boltzmann Machines}.
\newblock \bibinfo{journal}{\emph{IEEE Transactions on Neural Networks}}
  \bibinfo{volume}{21}, \bibinfo{number}{11} (\bibinfo{date}{Nov}
  \bibinfo{year}{2010}), \bibinfo{pages}{1780--1792}.
\newblock
\showISSN{1045-9227}


\bibitem[\protect\citeauthoryear{Merolla, Arthur, Alvarez-Icaza, Cassidy,
  Sawada, Akopyan, Jackson, Imam, Guo, Nakamura, et~al\mbox{.}}{Merolla
  et~al\mbox{.}}{2014}]%
        {Merolla2014}
\bibfield{author}{\bibinfo{person}{Paul~A Merolla}, \bibinfo{person}{John~V
  Arthur}, \bibinfo{person}{Rodrigo Alvarez-Icaza}, \bibinfo{person}{Andrew~S
  Cassidy}, \bibinfo{person}{Jun Sawada}, \bibinfo{person}{Filipp Akopyan},
  \bibinfo{person}{Bryan~L Jackson}, \bibinfo{person}{Nabil Imam},
  \bibinfo{person}{Chen Guo}, \bibinfo{person}{Yutaka Nakamura},
  {et~al\mbox{.}}} \bibinfo{year}{2014}\natexlab{}.
\newblock \showarticletitle{A million spiking-neuron integrated circuit with a
  scalable communication network and interface}.
\newblock \bibinfo{journal}{\emph{Science}} \bibinfo{volume}{345},
  \bibinfo{number}{6197} (\bibinfo{year}{2014}), \bibinfo{pages}{668--673}.
\newblock


\bibitem[\protect\citeauthoryear{Mizrahi, Hirtzlin, Fukushima, Kubota, Yuasa,
  Grollier, and Querlioz}{Mizrahi et~al\mbox{.}}{2018}]%
        {mizrahi2018neural}
\bibfield{author}{\bibinfo{person}{Alice Mizrahi}, \bibinfo{person}{Tifenn
  Hirtzlin}, \bibinfo{person}{Akio Fukushima}, \bibinfo{person}{Hitoshi
  Kubota}, \bibinfo{person}{Shinji Yuasa}, \bibinfo{person}{Julie Grollier},
  {and} \bibinfo{person}{Damien Querlioz}.} \bibinfo{year}{2018}\natexlab{}.
\newblock \showarticletitle{Neural-like computing with populations of
  superparamagnetic basis functions}.
\newblock \bibinfo{journal}{\emph{Nature communications}} \bibinfo{volume}{9},
  \bibinfo{number}{1} (\bibinfo{year}{2018}), \bibinfo{pages}{1533}.
\newblock


\bibitem[\protect\citeauthoryear{Ostwal, Debashis, Faria, Chen, and
  Appenzeller}{Ostwal et~al\mbox{.}}{2018}]%
        {Vaibhav2018}
\bibfield{author}{\bibinfo{person}{Vaibhav Ostwal},
  \bibinfo{person}{Punyashloka Debashis}, \bibinfo{person}{Rafatul Faria},
  \bibinfo{person}{Zhihong Chen}, {and} \bibinfo{person}{Joerg Appenzeller}.}
  \bibinfo{year}{2018}\natexlab{}.
\newblock \showarticletitle{Spin-torque devices with hard axis initialization
  as Stochastic Binary Neurons}.
\newblock \bibinfo{journal}{\emph{Scientific reports}} \bibinfo{volume}{8},
  \bibinfo{number}{1} (\bibinfo{year}{2018}), \bibinfo{pages}{16689}.
\newblock


\bibitem[\protect\citeauthoryear{Parkin, Kaiser, Panchula, Rice, Hughes,
  Samant, and Yang}{Parkin et~al\mbox{.}}{2004}]%
        {parkin2004}
\bibfield{author}{\bibinfo{person}{Stuart~SP Parkin},
  \bibinfo{person}{Christian Kaiser}, \bibinfo{person}{Alex Panchula},
  \bibinfo{person}{Philip~M Rice}, \bibinfo{person}{Brian Hughes},
  \bibinfo{person}{Mahesh Samant}, {and} \bibinfo{person}{See-Hun Yang}.}
  \bibinfo{year}{2004}\natexlab{}.
\newblock \showarticletitle{Giant tunnelling magnetoresistance at room
  temperature with MgO (100) tunnel barriers}.
\newblock \bibinfo{journal}{\emph{Nature materials}} \bibinfo{volume}{3},
  \bibinfo{number}{12} (\bibinfo{year}{2004}), \bibinfo{pages}{862}.
\newblock


\bibitem[\protect\citeauthoryear{Pufall, Rippard, Kaka, Russek, Silva, Katine,
  and Carey}{Pufall et~al\mbox{.}}{2004}]%
        {Pufall2004}
\bibfield{author}{\bibinfo{person}{M.~R. Pufall}, \bibinfo{person}{W.~H.
  Rippard}, \bibinfo{person}{Shehzaad Kaka}, \bibinfo{person}{S.~E. Russek},
  \bibinfo{person}{T.~J. Silva}, \bibinfo{person}{Jordan Katine}, {and}
  \bibinfo{person}{Matt Carey}.} \bibinfo{year}{2004}\natexlab{}.
\newblock \showarticletitle{Large-angle, gigahertz-rate random telegraph
  switching induced by spin-momentum transfer}.
\newblock \bibinfo{journal}{\emph{Phys. Rev. B}}  \bibinfo{volume}{69}
  (\bibinfo{date}{Jun} \bibinfo{year}{2004}), \bibinfo{pages}{214409}.
\newblock
Issue 21.
\urldef\tempurl%
\url{https://doi.org/10.1103/PhysRevB.69.214409}
\showDOI{\tempurl}


\bibitem[\protect\citeauthoryear{Roy, Sengupta, and Shim}{Roy
  et~al\mbox{.}}{2018}]%
        {roy2018}
\bibfield{author}{\bibinfo{person}{Kaushik Roy}, \bibinfo{person}{Abhronil
  Sengupta}, {and} \bibinfo{person}{Yong Shim}.}
  \bibinfo{year}{2018}\natexlab{}.
\newblock \showarticletitle{Perspective: Stochastic magnetic devices for
  cognitive computing}.
\newblock \bibinfo{journal}{\emph{Journal of Applied Physics}}
  \bibinfo{volume}{123}, \bibinfo{number}{21} (\bibinfo{year}{2018}),
  \bibinfo{pages}{210901}.
\newblock


\bibitem[\protect\citeauthoryear{Sankey, Cui, Sun, Slonczewski, Buhrman, and
  Ralph}{Sankey et~al\mbox{.}}{2008}]%
        {sankey2008}
\bibfield{author}{\bibinfo{person}{Jack~C Sankey}, \bibinfo{person}{Yong-Tao
  Cui}, \bibinfo{person}{Jonathan~Z Sun}, \bibinfo{person}{John~C Slonczewski},
  \bibinfo{person}{Robert~A Buhrman}, {and} \bibinfo{person}{Daniel~C Ralph}.}
  \bibinfo{year}{2008}\natexlab{}.
\newblock \showarticletitle{Measurement of the spin-transfer-torque vector in
  magnetic tunnel junctions}.
\newblock \bibinfo{journal}{\emph{Nature Physics}} \bibinfo{volume}{4},
  \bibinfo{number}{1} (\bibinfo{year}{2008}), \bibinfo{pages}{67}.
\newblock


\bibitem[\protect\citeauthoryear{Sarikaya, Hinton, and Deoras}{Sarikaya
  et~al\mbox{.}}{2014}]%
        {Sarikaya2014}
\bibfield{author}{\bibinfo{person}{R. Sarikaya}, \bibinfo{person}{G.~E.
  Hinton}, {and} \bibinfo{person}{A. Deoras}.} \bibinfo{year}{2014}\natexlab{}.
\newblock \showarticletitle{Application of Deep Belief Networks for Natural
  Language Understanding}.
\newblock \bibinfo{journal}{\emph{IEEE/ACM Transactions on Audio, Speech, and
  Language Processing}} \bibinfo{volume}{22}, \bibinfo{number}{4}
  (\bibinfo{date}{April} \bibinfo{year}{2014}), \bibinfo{pages}{778--784}.
\newblock
\showISSN{2329-9290}
\urldef\tempurl%
\url{https://doi.org/10.1109/TASLP.2014.2303296}
\showDOI{\tempurl}


\bibitem[\protect\citeauthoryear{Scott}{Scott}{1998}]%
        {scott1998}
\bibfield{author}{\bibinfo{person}{JF Scott}.} \bibinfo{year}{1998}\natexlab{}.
\newblock \showarticletitle{High-dielectric constant thin films for dynamic
  random access memories (DRAM)}.
\newblock \bibinfo{journal}{\emph{Annual review of materials science}}
  \bibinfo{volume}{28}, \bibinfo{number}{1} (\bibinfo{year}{1998}),
  \bibinfo{pages}{79--100}.
\newblock


\bibitem[\protect\citeauthoryear{Sengupta, Banerjee, and Roy}{Sengupta
  et~al\mbox{.}}{2016a}]%
        {Sengupta2016hybrid}
\bibfield{author}{\bibinfo{person}{Abhronil Sengupta},
  \bibinfo{person}{Aparajita Banerjee}, {and} \bibinfo{person}{Kaushik Roy}.}
  \bibinfo{year}{2016}\natexlab{a}.
\newblock \showarticletitle{Hybrid Spintronic-CMOS Spiking Neural Network with
  On-Chip Learning: Devices, Circuits, and Systems}.
\newblock \bibinfo{journal}{\emph{Phys. Rev. Applied}}  \bibinfo{volume}{6}
  (\bibinfo{date}{Dec} \bibinfo{year}{2016}), \bibinfo{pages}{064003}.
\newblock
Issue 6.


\bibitem[\protect\citeauthoryear{Sengupta, Panda, Wijesinghe, Kim, and
  Roy}{Sengupta et~al\mbox{.}}{2016b}]%
        {sengupta2016SREP}
\bibfield{author}{\bibinfo{person}{Abhronil Sengupta},
  \bibinfo{person}{Priyadarshini Panda}, \bibinfo{person}{Parami Wijesinghe},
  \bibinfo{person}{Yusung Kim}, {and} \bibinfo{person}{Kaushik Roy}.}
  \bibinfo{year}{2016}\natexlab{b}.
\newblock \showarticletitle{Magnetic tunnel junction mimics stochastic cortical
  spiking neurons}.
\newblock \bibinfo{journal}{\emph{Scientific reports}}  \bibinfo{volume}{6}
  (\bibinfo{year}{2016}), \bibinfo{pages}{30039}.
\newblock


\bibitem[\protect\citeauthoryear{Sengupta, Parsa, Han, and Roy}{Sengupta
  et~al\mbox{.}}{2016c}]%
        {Sengupta2016TED}
\bibfield{author}{\bibinfo{person}{A. Sengupta}, \bibinfo{person}{M. Parsa},
  \bibinfo{person}{B. Han}, {and} \bibinfo{person}{K. Roy}.}
  \bibinfo{year}{2016}\natexlab{c}.
\newblock \showarticletitle{Probabilistic Deep Spiking Neural Systems Enabled
  by Magnetic Tunnel Junction}.
\newblock \bibinfo{journal}{\emph{IEEE Transactions on Electron Devices}}
  \bibinfo{volume}{63}, \bibinfo{number}{7} (\bibinfo{date}{July}
  \bibinfo{year}{2016}), \bibinfo{pages}{2963--2970}.
\newblock
\showISSN{0018-9383}
\urldef\tempurl%
\url{https://doi.org/10.1109/TED.2016.2568762}
\showDOI{\tempurl}


\bibitem[\protect\citeauthoryear{Sheri, Rafique, Pedrycz, and Jeon}{Sheri
  et~al\mbox{.}}{2015}]%
        {SHERI2015}
\bibfield{author}{\bibinfo{person}{Ahmad~Muqeem Sheri}, \bibinfo{person}{Aasim
  Rafique}, \bibinfo{person}{Witold Pedrycz}, {and} \bibinfo{person}{Moongu
  Jeon}.} \bibinfo{year}{2015}\natexlab{}.
\newblock \showarticletitle{Contrastive divergence for memristor-based
  restricted Boltzmann machine}.
\newblock \bibinfo{journal}{\emph{Engineering Applications of Artificial
  Intelligence}}  \bibinfo{volume}{37} (\bibinfo{year}{2015}),
  \bibinfo{pages}{336 -- 342}.
\newblock
\showISSN{0952-1976}


\bibitem[\protect\citeauthoryear{Stengel and Spaldin}{Stengel and
  Spaldin}{2006}]%
        {stengel2006}
\bibfield{author}{\bibinfo{person}{Massimiliano Stengel} {and}
  \bibinfo{person}{Nicola~A Spaldin}.} \bibinfo{year}{2006}\natexlab{}.
\newblock \showarticletitle{Origin of the dielectric dead layer in nanoscale
  capacitors}.
\newblock \bibinfo{journal}{\emph{Nature}} \bibinfo{volume}{443},
  \bibinfo{number}{7112} (\bibinfo{year}{2006}), \bibinfo{pages}{679}.
\newblock


\bibitem[\protect\citeauthoryear{Strukov, Snider, Stewart, and
  Williams}{Strukov et~al\mbox{.}}{2008}]%
        {strukov2008}
\bibfield{author}{\bibinfo{person}{Dmitri~B Strukov},
  \bibinfo{person}{Gregory~S Snider}, \bibinfo{person}{Duncan~R Stewart}, {and}
  \bibinfo{person}{R~Stanley Williams}.} \bibinfo{year}{2008}\natexlab{}.
\newblock \showarticletitle{The missing memristor found}.
\newblock \bibinfo{journal}{\emph{nature}} \bibinfo{volume}{453},
  \bibinfo{number}{7191} (\bibinfo{year}{2008}), \bibinfo{pages}{80}.
\newblock


\bibitem[\protect\citeauthoryear{Sutton, Camsari, Behin-Aein, and Datta}{Sutton
  et~al\mbox{.}}{2017}]%
        {sutton2017intrinsic}
\bibfield{author}{\bibinfo{person}{Brian Sutton}, \bibinfo{person}{Kerem~Yunus
  Camsari}, \bibinfo{person}{Behtash Behin-Aein}, {and}
  \bibinfo{person}{Supriyo Datta}.} \bibinfo{year}{2017}\natexlab{}.
\newblock \showarticletitle{Intrinsic optimization using stochastic
  nanomagnets}.
\newblock \bibinfo{journal}{\emph{Scientific reports}}  \bibinfo{volume}{7}
  (\bibinfo{year}{2017}), \bibinfo{pages}{44370}.
\newblock


\bibitem[\protect\citeauthoryear{Tanaka and Okutomi}{Tanaka and
  Okutomi}{2014}]%
        {Tanaka2014}
\bibfield{author}{\bibinfo{person}{M. Tanaka} {and} \bibinfo{person}{M.
  Okutomi}.} \bibinfo{year}{2014}\natexlab{}.
\newblock \showarticletitle{A Novel Inference of a Restricted Boltzmann
  Machine}. In \bibinfo{booktitle}{\emph{2014 22nd International Conference on
  Pattern Recognition}}. \bibinfo{pages}{1526--1531}.
\newblock
\showISSN{1051-4651}
\urldef\tempurl%
\url{https://doi.org/10.1109/ICPR.2014.271}
\showDOI{\tempurl}


\bibitem[\protect\citeauthoryear{Wang, Cai, Cao, Zhou, Wrona, Peng, Yang, Wei,
  Kang, Zhang, et~al\mbox{.}}{Wang et~al\mbox{.}}{2018}]%
        {wang2018}
\bibfield{author}{\bibinfo{person}{Mengxing Wang}, \bibinfo{person}{Wenlong
  Cai}, \bibinfo{person}{Kaihua Cao}, \bibinfo{person}{Jiaqi Zhou},
  \bibinfo{person}{Jerzy Wrona}, \bibinfo{person}{Shouzhong Peng},
  \bibinfo{person}{Huaiwen Yang}, \bibinfo{person}{Jiaqi Wei},
  \bibinfo{person}{Wang Kang}, \bibinfo{person}{Youguang Zhang},
  {et~al\mbox{.}}} \bibinfo{year}{2018}\natexlab{}.
\newblock \showarticletitle{Current-induced magnetization switching in
  atom-thick tungsten engineered perpendicular magnetic tunnel junctions with
  large tunnel magnetoresistance}.
\newblock \bibinfo{journal}{\emph{Nature communications}} \bibinfo{volume}{9},
  \bibinfo{number}{1} (\bibinfo{year}{2018}), \bibinfo{pages}{671}.
\newblock


\bibitem[\protect\citeauthoryear{Wang, Sukegawa, Shan, Mitani, and
  Inomata}{Wang et~al\mbox{.}}{2009}]%
        {wang2009}
\bibfield{author}{\bibinfo{person}{Wenhong Wang}, \bibinfo{person}{Hiroaki
  Sukegawa}, \bibinfo{person}{Rong Shan}, \bibinfo{person}{Seiji Mitani}, {and}
  \bibinfo{person}{Koichiro Inomata}.} \bibinfo{year}{2009}\natexlab{}.
\newblock \showarticletitle{Giant tunneling magnetoresistance up to 330\% at
  room temperature in sputter deposited Co 2 FeAl/MgO/CoFe magnetic tunnel
  junctions}.
\newblock \bibinfo{journal}{\emph{Applied Physics Letters}}
  \bibinfo{volume}{95}, \bibinfo{number}{18} (\bibinfo{year}{2009}),
  \bibinfo{pages}{182502}.
\newblock


\bibitem[\protect\citeauthoryear{Yuan and Parhi}{Yuan and Parhi}{2017}]%
        {Yuan2017}
\bibfield{author}{\bibinfo{person}{Bo Yuan} {and} \bibinfo{person}{Keshab~K.
  Parhi}.} \bibinfo{year}{2017}\natexlab{}.
\newblock \showarticletitle{VLSI Architectures for the Restricted Boltzmann
  Machine}.
\newblock \bibinfo{journal}{\emph{J. Emerg. Technol. Comput. Syst.}}
  \bibinfo{volume}{13}, \bibinfo{number}{3}, Article \bibinfo{articleno}{35}
  (\bibinfo{date}{May} \bibinfo{year}{2017}), \bibinfo{numpages}{19}~pages.
\newblock
\showISSN{1550-4832}
\urldef\tempurl%
\url{https://doi.org/10.1145/3007193}
\showDOI{\tempurl}


\bibitem[\protect\citeauthoryear{Yuasa, Nagahama, Fukushima, Suzuki, and
  Ando}{Yuasa et~al\mbox{.}}{2004}]%
        {yuasa2004giant}
\bibfield{author}{\bibinfo{person}{Shinji Yuasa}, \bibinfo{person}{Taro
  Nagahama}, \bibinfo{person}{Akio Fukushima}, \bibinfo{person}{Yoshishige
  Suzuki}, {and} \bibinfo{person}{Koji Ando}.} \bibinfo{year}{2004}\natexlab{}.
\newblock \showarticletitle{Giant room-temperature magnetoresistance in
  single-crystal Fe/MgO/Fe magnetic tunnel junctions.}
\newblock \bibinfo{journal}{\emph{Nature materials}} \bibinfo{volume}{3},
  \bibinfo{number}{12} (\bibinfo{year}{2004}).
\newblock


\bibitem[\protect\citeauthoryear{Zand, Camsari, Pyle, Ahmed, Kim, and
  DeMara}{Zand et~al\mbox{.}}{2018}]%
        {zandRDBN}
\bibfield{author}{\bibinfo{person}{Ramtin Zand}, \bibinfo{person}{Kerem~Yunus
  Camsari}, \bibinfo{person}{Steven~D. Pyle}, \bibinfo{person}{Ibrahim Ahmed},
  \bibinfo{person}{Chris~H. Kim}, {and} \bibinfo{person}{Ronald~F. DeMara}.}
  \bibinfo{year}{2018}\natexlab{}.
\newblock \showarticletitle{Low-Energy Deep Belief Networks Using Intrinsic
  Sigmoidal Spintronic-based Probabilistic Neurons}. In
  \bibinfo{booktitle}{\emph{Proceedings of the 2018 on Great Lakes Symposium on
  VLSI}} \emph{(\bibinfo{series}{GLSVLSI '18})}. \bibinfo{publisher}{ACM},
  \bibinfo{address}{Chicago, IL, USA}, \bibinfo{pages}{15--20}.
\newblock
\showISBNx{978-1-4503-5724-1}
\urldef\tempurl%
\url{https://doi.org/10.1145/3194554.3194558}
\showDOI{\tempurl}


\end{thebibliography}

\end{document}